# Analytical Gradients for Nuclear-Electronic Orbital Time-Dependent Density Functional Theory: Excited State Geometry Optimizations and Adiabatic Excitation Energies

Zhen Tao, Saswata Roy, Patrick E. Schneider, Fabijan Pavošević, and Sharon Hammes-Schiffer*

Department of Chemistry, Yale University
225 Prospect Street, New Haven, Connecticut 06520 USA
e-mail: sharon.hammes-schiffer@yale.edu

**Abstract**

The computational investigation of photochemical processes often entails the calculation of excited state geometries, energies, and energy gradients. The nuclear-electronic orbital (NEO) approach treats specified nuclei, typically protons, quantum mechanically on the same level as the electrons, thereby including the associated nuclear quantum effects and non-Born-Oppenheimer behavior into quantum chemistry calculations. The multicomponent density functional theory (NEO-DFT) and time-dependent DFT (NEO-TDDFT) methods allow efficient calculations of ground and excited states, respectively. Herein, the analytical gradients are derived and implemented for the NEO-TDDFT method and the associated Tamm-Dancoff approximation (NEO-TDA). The programmable equations for these analytical gradients, as well as the NEO-DFT analytical Hessian, are provided. The NEO approach includes the anharmonic zero-point energy and density delocalization associated with the quantum protons, as well as vibronic mixing, in geometry optimizations and energy calculations of ground and excited states. The harmonic zero-point energy associated with the other nuclei can be computed via the NEO Hessian. This approach is used to compute the 0-0 adiabatic excitation energies for a set of nine small molecules with all protons quantized, exhibiting slight improvement over the conventional electronic approach. Geometry optimizations of two excited state intramolecular proton transfer systems, [2,2′-bipyridyl]-3-ol and [2,2′-bipyridyl]-3,3′-diol, are performed with one and two quantized protons, respectively. The NEO calculations for these systems produce electronically excited state geometries with stronger intramolecular hydrogen bonds and similar relative stabilities compared to conventional electronic methods. This work provides the foundation for nonadiabatic dynamics simulations of fundamental processes such as photoinduced proton transfer and proton-coupled electron transfer.

## 1. Introduction

Within the Born-Oppenheimer approximation, nuclei move on a potential energy surface according to forces determined from the gradient of the energy with respect to the nuclear coordinates. The implementation of analytical gradients for excited state potential energy surfaces allows excited state geometry optimizations,[1] calculation of adiabatic excitation energies,[2] and nonadiabatic dynamics on adiabatic surfaces.[3-4] Among the various excited state electronic structure methods, time-dependent density functional theory (TDDFT) provides a balance of accuracy and efficiency[5] and has been used extensively for trajectory-based nonadiabatic dynamics methods.[3, 6] Although TDDFT has well-known shortcomings in terms of describing excited states with charge transfer character[7-13] and conical intersections between the ground and first excited electronic states,[14] long-range corrected functionals,[15] the Tamm-Dancoff approximation (TDA),[16] and approximate nonadiabatic dynamics schemes can address some of these issues in certain cases.[17-19] Thus, the TDDFT and TDDFT-TDA approaches have provided insights into a wide range of photochemical processes.

Typically nuclear quantum effects such as zero-point energy (ZPE) are included as corrections to the adiabatic potential energy surfaces when computing adiabatic excitation energies and are not included at all in excited state geometry optimizations. Moreover, calculations of excited state geometries and energies on adiabatic potential energy surfaces often neglect non-Born-Oppenheimer effects. However these effects are important for studying many fundamental phenomena, such as hydrogen tunneling and proton-coupled electron transfer.[20-21] An efficient approach for including nuclear quantum effects and non-Born-Oppenheimer behavior into excited state calculations is the nuclear-electronic orbital (NEO) method.[22-23] The NEO Hamiltonian depends explicitly on the coordinates of specified nuclei as well as all electrons, and a mixed nuclear-electronic Schrödinger equation is solved with molecular orbital techniques. As a result, nuclear quantum effects such as density delocalization and ZPE are inherently included for the specified nuclei, and non-Born-Oppenheimer effects between the electrons and these nuclei are captured. Both wave function-based methods and density functional theory (DFT) methods have been developed within the NEO framework to compute ground[24-32] and excited state[33-36] properties. These NEO methods belong to the more general class of multicomponent quantum chemistry methods that treat different types of particles quantum mechanically on the same level.[37-44] Although multicomponent wave function based methods have the advantage of allowing systematic improvement, multicomponent DFT enables the study of larger systems due to its lower computational scaling. The NEO approach is particularly computationally efficient because only specified nuclei, typically protons, are treated quantum mechanically on the same level as the electrons.

The NEO-DFT method,[24, 45] in conjunction with suitable electron-proton correlation functionals,[25-28] has been shown to provide accurate proton densities, vibrationally averaged geometries, and proton affinities. The linear response NEO-TDDFT method[33] allows the calculation of excited state properties, such as excitation energies, transition densities, and transition dipole moments. Within the adiabatic approximation, NEO-TDDFT can provide excitations that are predominantly electronic or protonic (i.e., vibrational) or have mixed electronic-protonic (i.e., vibronic) character corresponding to linear combinations of single electronic and protonic excitations. NEO-TDDFT is more computationally affordable than wave function-based methods such as the NEO equation-of-motion coupled cluster singles and doubles (NEO-EOM-CCSD) method[34, 36] and has been shown to be accurate for describing electronic and fundamental vibrational excitations. In particular, NEO-TDDFT produces fundamental proton vibrational excitation energies that are typically within 30 $cm^{-1}$ of grid-based benchmarks.[35] The lower-lying electronic excitation energies are similar to conventional TDDFT excitation energies,[33] but the higher electronic excitation energies can be shifted due to vibronic mixing. The NEO-TDDFT-TDA method significantly overestimates the proton vibrational frequencies but provides similar electronic excitation energies as the NEO-TDDFT method.[33, 35]

Herein, we derive and implement the NEO-TDDFT and NEO-TDA analytical gradients. These analytical gradients allow us to perform excited state geometry optimizations and compute adiabatic excitation energies, $\Delta E^{\mathrm{adia}}$, which are defined to be the energy difference between geometries optimized on the ground and excited states. Using analytical Hessians for the NEO-DFT ground state and semi-numerical Hessians for the excited states, we are also able to calculate the 0-0 adiabatic excitation energies, $\Delta E^{0-0}$, which include the ZPE corrections for the ground and excited state optimized geometries. This property can be compared directly to experimentally measured 0-0 adiabatic excitation energies.[2] The adiabatic and 0-0 adiabatic excitation energies are depicted in Figure 1a.

To benchmark the NEO methods, we calculate the 0-0 adiabatic excitation energies of select singlet, doublet, and triplet excited states for the nine small molecules shown in Figure 1b. All protons are treated quantum mechanically with NEO-TDDFT or NEO-TDA, and the resulting energies are compared to the conventional electronic TDDFT or TDA counterparts, as well as to experimental results. The ZPEs and associated anharmonicities of the quantum protons are inherently included in both the ground and excited state NEO energies, and the ZPE contributions from the classical nuclei are computed within the harmonic approximation for the NEO vibronic surfaces using the NEO Hessian. An advantage of the NEO approach over the conventional approach is the inclusion of nuclear quantum effects of the quantum protons during the geometry optimizations and the inclusion of anharmonicity in the ZPEs.

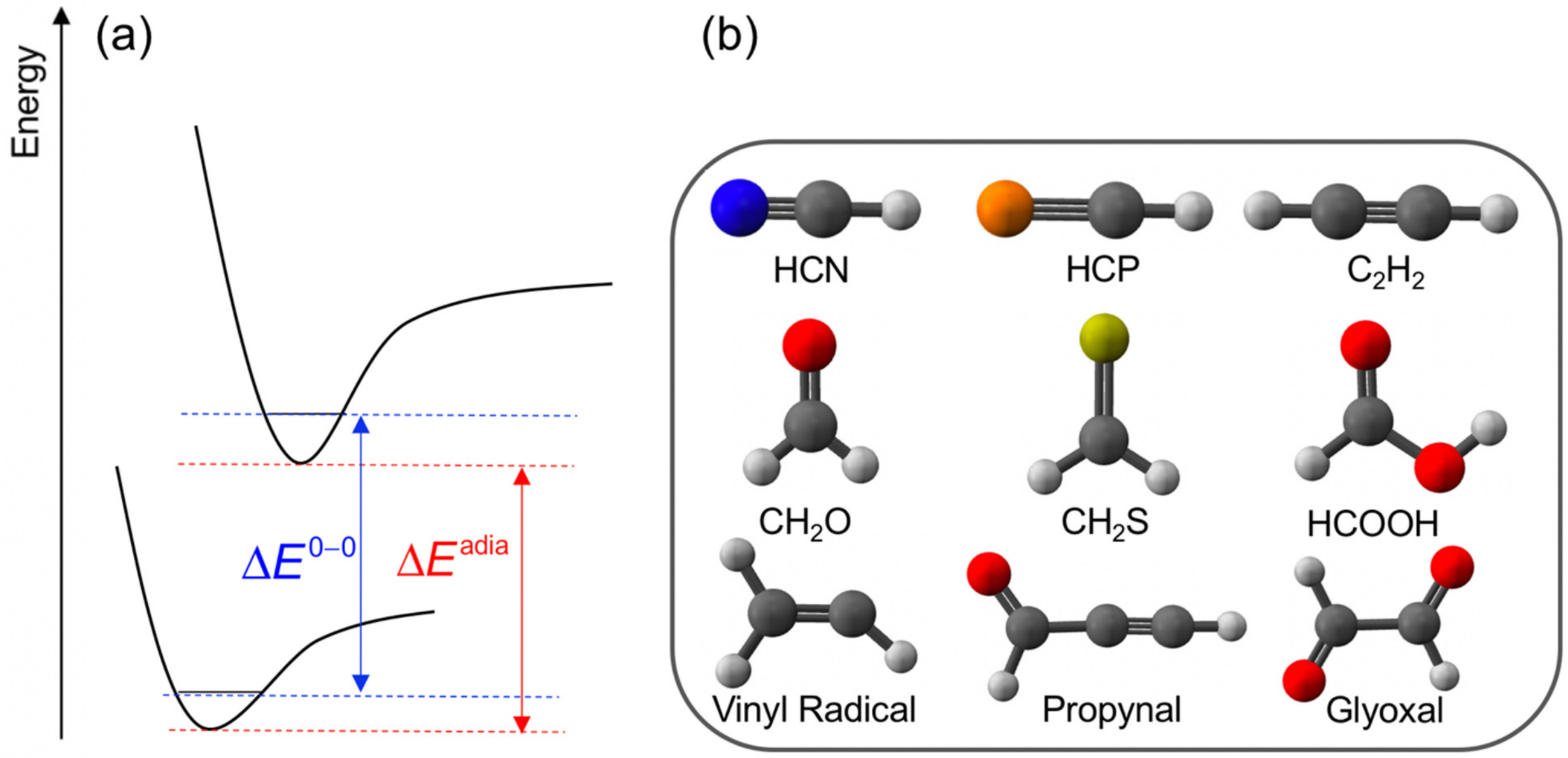

**Figure 1**. Schematic depiction of (a) adiabatic and 0-0 adiabatic excitation energies, $\Delta E^{\text{adia}}$ and $\Delta E^{0\text{–}0}$, respectively, and (b) the nine small molecules studied in this work. The adiabatic excitation energy $\Delta E^{\text{adia}}$ (red) is defined to be the energy difference between the optimized excited state and ground state geometries, and the 0-0 adiabatic excitation energy $\Delta E^{0\text{–}0}$ (blue) includes the ZPE contributions. The geometries given in (b) are NEO-DFT ground state optimized geometries.

In addition, we conduct excited state geometry optimizations of two prototype molecules for excited-state intramolecular proton transfer (ESIPT)[46] within the NEO framework to investigate the impact of nuclear quantum effects on the excited state geometries and energies. ESIPT molecules have a wide range of important applications in biological and chemical sciences due to desirable features such as large Stokes-shifted fluorescence emission and tunable photophysical characteristics.[46] The specific systems studied herein are [2,2'-bipyridyl]-3-ol, denoted BPOH, and [2,2'-bipyridyl]-3,3'-diol, denoted $BP(OH)_2$, as depicted in Figure 2. Both of these molecules undergo ESIPT in the $\pi\pi^*$ electronically excited state, corresponding to isomerization from the enol (E) to the keto (K) form. Upon photoexcitation of BPOH, single proton transfer produces the keto form, K*, which is observed to emit fluorescence.[47-48] However, twisted intramolecular charge transfer (TICT) after ESIPT leads to the twisted keto form, Kt*, which is proposed to be responsible for fluorescence quenching in this molecule.[46-47, 49-50] In contrast, experimental[51-52] and computational[53-57] data for $BP(OH)_2$ implicate a branched process involving both a concerted double proton transfer mechanism directly to the double keto form, KK*, and a stepwise mechanism with a single proton transfer intermediate, EK*. Herein, our goal is to compute the excited state optimized geometries and their relative stabilities for these two systems with NEO-TDDFT. By treating the transferring proton(s) quantum mechanically, nuclear quantum effects such as zero-point energy and density delocalization, as well as anharmonicity, are included inherently in the excited state geometries and excitation energies.

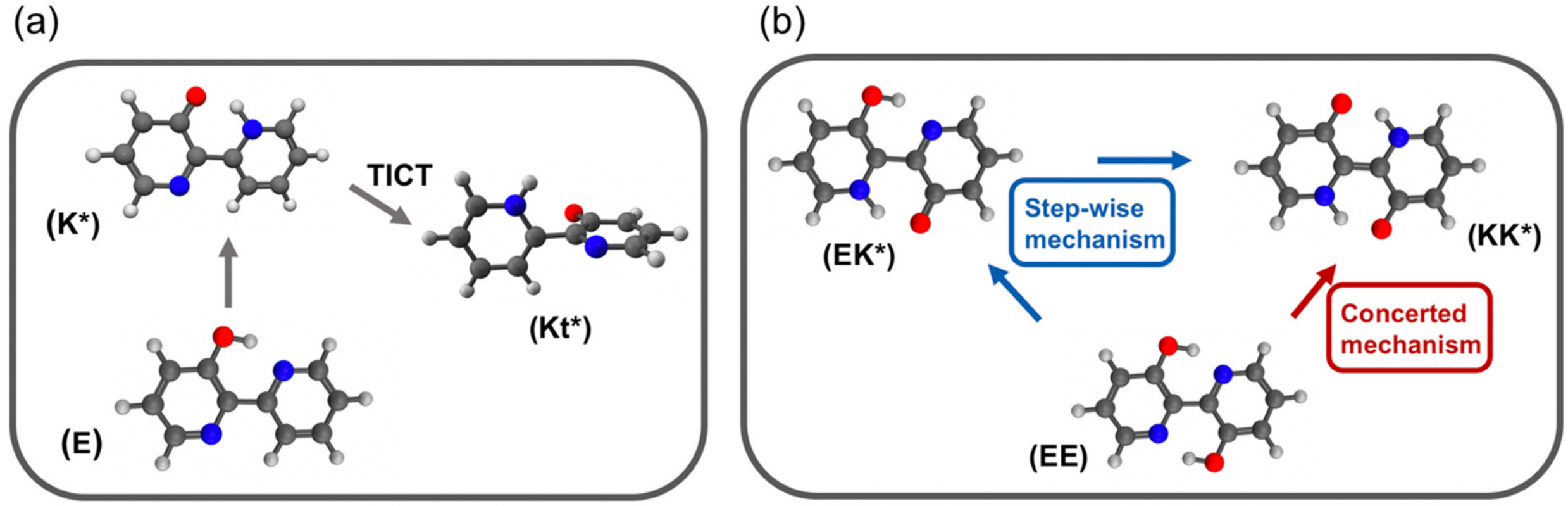

**Figure 2**. Schematic depictions of select ground state and excited state structures of (a) BPOH and (b) $BP(OH)_2$ . The ground states of both molecules are in the enol (E or EE) form, and the excited states are denoted by *. Photoexcitation to the excited electronic state (E* or EE*) induces proton transfer from the oxygen to the nitrogen, transforming the enol tautomer to the excited state keto (K*) tautomer. The TICT reaction in BPOH leads to the twisted keto form Kt*, which is proposed to quench fluorescence. The $BP(OH)_2$ system exhibits both concerted double proton transfer directly to KK* and stepwise double proton transfer with EK* as an intermediate.

An outline of this paper is as follows. Section 2 presents the derivation of the NEO-TDDFT and NEO-TDA analytical gradients, with additional details provided in the Supporting Information. Section 3 presents NEO-TDDFT and NEO-TDA calculations of the 0-0 adiabatic excitation energies for a set of nine molecules and NEO-TDDFT excited state geometry optimizations and energies for the BPOH and $BP(OH)_2$ systems. Section 4 provides concluding remarks. This work provides the foundation for a wide range of applications, including on-the-fly dynamics on adiabatic excited state electron-proton vibronic surfaces and nonadiabatic dynamics on these surfaces.

## 2. Theory

### 2.1 NEO-DFT Kohn-Sham Formalism

In NEO-DFT, the total energy is a functional of the one-particle electronic and protonic densities $\rho^{\mathrm{e}}$ and $\rho^{\mathrm{p}}$, respectively. This energy functional can be written as

$$E[\rho^{\mathrm{e}},\rho^{\mathrm{p}}]=E^{\mathrm{ext}}[\rho^{\mathrm{e}},\rho^{\mathrm{p}}]+E^{\mathrm{ref}}[\rho^{\mathrm{e}},\rho^{\mathrm{p}}]+E^{\mathrm{exc}}[\rho^{\mathrm{e}}]+E^{\mathrm{pxc}}[\rho^{\mathrm{p}}]+E^{\mathrm{epc}}[\rho^{\mathrm{e}},\rho^{\mathrm{p}}] \tag{1}$$

where $E^{\mathrm{ext}}[\rho^{\mathrm{e}},\rho^{\mathrm{p}}]$ is the interaction of the two densities with the external potential created by the classical nuclei, and $E^{\mathrm{ref}}[\rho^{\mathrm{e}},\rho^{\mathrm{p}}]$ is the sum of the noninteracting kinetic energies of the electrons and quantum protons and the electron-electron, proton-proton, and electron-proton classical Coulomb energies. Moreover, $E^{\mathrm{exc}}[\rho^{\mathrm{e}}]$, $E^{\mathrm{pxc}}[\rho^{\mathrm{p}}]$, and $E^{\mathrm{epc}}[\rho^{\mathrm{e}},\rho^{\mathrm{p}}]$ are the electron-electron exchange-correlation (exc) functional, the proton-proton exchange-correlation (pxc) functional, and the electron-proton correlation (epc) functional. Similar to the exc functionals, the latter two functionals also need to be approximated. In molecular systems, the pxc energy is generally small due to the localized nature of proton orbitals.[23, 58] Hence, we use the exact exchange for the pxc functional to remove the proton-proton self-interaction terms. Unlike exc functionals, which have been widely explored in the community, the choices of epc functionals are rather limited, despite its importance in obtaining even qualitatively correct results. Our group has developed a series of epc functionals[25-28] based on the Colle-Salvetti formalism[59] and has demonstrated their capabilities in computing accurate proton and deuterium densities, proton affinities, and vibrationally averaged geometries.

In the multicomponent Kohn-Sham formalism, the reference system is the product of electronic and protonic Slater determinants $\Phi^{\mathrm{e}}(\mathbf{x}^{\mathrm{e}})$ and $\Phi^{\mathrm{p}}(\mathbf{x}^{\mathrm{p}})$, respectively, where $\mathbf{x}^{\mathrm{e}}$ and $\mathbf{x}^{\mathrm{p}}$ are the collective spatial and spin coordinates of the electrons and quantum protons, respectively. The electronic and protonic Slater determinants are composed of orthogonal electronic and protonic orbitals, $\psi_i^{\mathrm{e}}$ and $\psi_{i'}^{\mathrm{p}}$ , respectively, which are assumed to be real for simplicity. These orbitals are further expanded in electronic and protonic basis functions $\phi_\mu^{\mathrm{e}}$ and $\phi_{\mu'}^{\mathrm{p}}$, respectively, with orbital coefficients given by matrices $\mathbf{C}^{\mathrm{e}}$ and $\mathbf{C}^{\mathrm{p}}$, which define the density matrices $\mathbf{D}^{\mathrm{e}}$ and $\mathbf{D}^{\mathrm{p}}$, respectively. Using this notation, the electronic and protonic densities can be expressed as

$$\begin{aligned}
\rho_\sigma^{\mathrm{e}} &= \sum_i^{N_\sigma^{\mathrm{e}}} |\psi_{i\sigma}^{\mathrm{e}}|^2 = \sum_{\mu\nu}\left(\sum_i^{N_\sigma^{\mathrm{e}}} C_{\mu i\sigma}^{\mathrm{e}} C_{\nu i\sigma}^{\mathrm{e}}\right)\phi_\mu^{\mathrm{e}}\phi_\nu^{\mathrm{e}} = \sum_{\mu\nu} D_{\mu\nu\sigma}^{\mathrm{e}}\phi_\mu^{\mathrm{e}}\phi_\nu^{\mathrm{e}} \\
\rho^{\mathrm{p}} &= \sum_{i'}^{N^{\mathrm{p}}} |\psi_{i'}^{\mathrm{p}}|^2 = \sum_{\mu'\nu'}\left(\sum_{i'}^{N^{\mathrm{p}}} C_{\mu'i'}^{\mathrm{p}} C_{\nu'i'}^{\mathrm{p}}\right)\phi_{\mu'}^{\mathrm{p}}\phi_{\nu'}^{\mathrm{p}} = \sum_{\mu'\nu'} D_{\mu'\nu'}^{\mathrm{p}}\phi_{\mu'}^{\mathrm{p}}\phi_{\nu'}^{\mathrm{p}}
\end{aligned} \tag{2}$$

where $N^{\mathrm{p}}$ is the number of quantum protons, and $N^{\mathrm{e}}$ is the number of electrons with a subscript σ indicating the electron spin. Throughout this paper, the index σ and its primed versions are used exclusively to denote electron spin. All protons are treated as high spin, and therefore no index is required for the proton spin. The indices $p,q,r,s...$ denote general electronic spatial orbitals, $i,j,k,l...$ denote occupied orbitals, and $a,b,c,d...$ denote unoccupied orbitals. The indices $\mu,\nu,\kappa,\lambda...$ denote electronic basis functions. The primed versions correspond to the protonic counterparts.

The electronic and protonic density matrices can be obtained from the following NEO Kohn-Sham equations

$$\begin{aligned}\mathbf{F}^{\mathrm{e}}\mathbf{C}^{\mathrm{e}} &= \mathbf{S}^{\mathrm{e}}\mathbf{C}^{\mathrm{e}}\boldsymbol{\varepsilon}^{\mathrm{e}} \\ \mathbf{F}^{\mathrm{p}}\mathbf{C}^{\mathrm{p}} &= \mathbf{S}^{\mathrm{p}}\mathbf{C}^{\mathrm{p}}\boldsymbol{\varepsilon}^{\mathrm{p}}\end{aligned} \tag{3}$$

Here the electronic and protonic Kohn-Sham matrices $\mathbf{F}^{\mathrm{e}}$ and $\mathbf{F}^{\mathrm{p}}$ and overlap matrices $\mathbf{S}^{\mathrm{e}}$ and $\mathbf{S}^{\mathrm{p}}$ are defined in terms of the basis functions $\phi_{\mu}^{\mathrm{e}}$ and $\phi_{\mu'}^{\mathrm{p}}$, and $\boldsymbol{\varepsilon}^{\mathrm{e}}$ and $\boldsymbol{\varepsilon}^{\mathrm{p}}$ are the orbital energy matrices. The electronic and protonic Kohn-Sham matrices are defined in the orbital basis as

$$\begin{aligned}F_{pq\sigma}^{\mathrm{e}} &= h_{pq\sigma}^{\mathrm{e}} + \sum_{i\sigma'}\left[\left(pq\sigma \mid ii\sigma'\right) - c_{\mathrm{x}}\delta_{\sigma\sigma'}\left(pi\sigma \mid iq\sigma\right)\right] - \sum_{i'}\left(pq\sigma \mid i'i'\right) + v_{pq\sigma}^{\mathrm{exc}} + v_{pq\sigma}^{\mathrm{epc,e}} \\ F_{p'q'}^{\mathrm{p}} &= h_{p'q'}^{\mathrm{p}} + \sum_{i'}\left[\left(p'q' \mid i'i'\right) - \left(p'i' \mid i'q'\right)\right] - \sum_{i\sigma}\left(p'q' \mid ii\sigma\right) + v_{p'q'}^{\mathrm{epc,p}}\end{aligned} \tag{4}$$

Here $h^{\mathrm{e}}$ and $h^{\mathrm{p}}$ are the electronic and protonic core Hamiltonians, respectively. The electron-electron integrals $(pq\sigma \mid rs\sigma')$ are defined as $\int d\mathbf{r}_1^{\mathrm{e}}\int d\mathbf{r}_2^{\mathrm{e}}\psi_{p\sigma}^{\mathrm{e}}(\mathbf{r}_1^{\mathrm{e}})\psi_{q\sigma}^{\mathrm{e}}(\mathbf{r}_1^{\mathrm{e}})\frac{1}{\left|\mathbf{r}_1^{\mathrm{e}} - \mathbf{r}_2^{\mathrm{e}}\right|}\psi_{r\sigma'}^{\mathrm{e}}(\mathbf{r}_2^{\mathrm{e}})\psi_{s\sigma'}^{\mathrm{e}}(\mathbf{r}_2^{\mathrm{e}})$. The proton-proton and electron-proton integrals are defined analogously. The epc potentials are defined as $v_{\sigma}^{\mathrm{epc,e}} = \frac{\delta E^{\mathrm{epc}}}{\delta\rho_{\sigma}^{\mathrm{e}}}$ and $v^{\mathrm{epc,p}} = \frac{\delta E^{\mathrm{epc}}}{\delta\rho^{\mathrm{p}}}$. The exc functional used in Eq. (4) is a hybrid functional, where $c_{\mathrm{x}}$ is the parameter for the amount of Hartree-Fock exact exchange in the hybrid functional, and $v_{\sigma}^{\mathrm{exc}} = \frac{\delta E^{\mathrm{exc}}}{\delta\rho_{\sigma}^{\mathrm{e}}}$. Note that the Kohn-Sham matrices in Eqs. (3) and (4) are related through a basis transformation (often called an atomic orbital to molecular orbital transformation for electronic orbitals) using the orbital coefficient matrices. Moreover, $F_{pq\sigma}^{\mathrm{e}} = \delta_{pq}\varepsilon_{p\sigma}^{\mathrm{e}}$ and $F_{p'q'}^{\mathrm{p}} = \delta_{p'q'}\varepsilon_{p'}^{\mathrm{p}}$ when the coefficient matrices are NEO-DFT solutions and diagonalize the Kohn-Sham matrix.

## 2.2 NEO-TDDFT Lagrangian

The NEO-TDDFT equations have been derived from the linear response theory that describes the changes in the density matrices induced by small perturbations to the potentials. The detailed derivation can be found in the Supporting Information of Ref. [33]. Here we start directly from the working equation of NEO-TDDFT,[33]

$$\begin{pmatrix} \mathbf{A}^{\mathrm{e}} & \mathbf{B}^{\mathrm{e}} & \mathbf{C} & \mathbf{C} \\ \mathbf{B}^{\mathrm{e}} & \mathbf{A}^{\mathrm{e}} & \mathbf{C} & \mathbf{C} \\ \mathbf{C}^{\mathrm{T}} & \mathbf{C}^{\mathrm{T}} & \mathbf{A}^{\mathrm{p}} & \mathbf{B}^{\mathrm{p}} \\ \mathbf{C}^{\mathrm{T}} & \mathbf{C}^{\mathrm{T}} & \mathbf{B}^{\mathrm{p}} & \mathbf{A}^{\mathrm{p}} \end{pmatrix} \begin{pmatrix} \mathbf{X}^{\mathrm{e}} \\ \mathbf{Y}^{\mathrm{e}} \\ \mathbf{X}^{\mathrm{p}} \\ \mathbf{Y}^{\mathrm{p}} \end{pmatrix} = \omega \begin{pmatrix} \mathbf{I} & 0 & 0 & 0 \\ 0 & -\mathbf{I} & 0 & 0 \\ 0 & 0 & \mathbf{I} & 0 \\ 0 & 0 & 0 & -\mathbf{I} \end{pmatrix} \begin{pmatrix} \mathbf{X}^{\mathrm{e}} \\ \mathbf{Y}^{\mathrm{e}} \\ \mathbf{X}^{\mathrm{p}} \\ \mathbf{Y}^{\mathrm{p}} \end{pmatrix} \tag{5}$$

The electronic block composed of $\mathbf{A}^{\mathrm{e}}$ and $\mathbf{B}^{\mathrm{e}}$ contains the response of the internal electronic potential to changes in the electronic density matrix due to the initial perturbation. The protonic block has the analogous composition. The coupling block composed of $\mathbf{C}$ and $\mathbf{C}^{\mathrm{T}}$ contains the response of the internal potential of one type of particle to the changes in the density matrix of the other type of particle. The specific definitions of the elements of these matrices are given below. The solutions of the NEO-TDDFT equation give the excitation energies $\omega$, the excitation amplitudes $\mathbf{X}^{\mathrm{e}}$ and $\mathbf{X}^{\mathrm{p}}$, and the de-excitation amplitudes $\mathbf{Y}^{\mathrm{e}}$ and $\mathbf{Y}^{\mathrm{p}}$.

For convenience, Eq.(5) can be reorganized into a structure that resembles the conventional TDDFT working equation[11]

$$\begin{pmatrix} \mathbf{A} & \mathbf{B} \\ \mathbf{B} & \mathbf{A} \end{pmatrix} \begin{pmatrix} \mathbf{X} \\ \mathbf{Y} \end{pmatrix} = \omega \begin{pmatrix} \mathbf{I} & 0 \\ 0 & -\mathbf{I} \end{pmatrix} \begin{pmatrix} \mathbf{X} \\ \mathbf{Y} \end{pmatrix} \tag{6}$$

where $\mathbf{A} = \begin{pmatrix} \mathbf{A}^{\mathrm{p}} & \mathbf{C}^{\mathrm{T}} \\ \mathbf{C} & \mathbf{A}^{\mathrm{e}} \end{pmatrix}$ and $\mathbf{B} = \begin{pmatrix} \mathbf{B}^{\mathrm{p}} & \mathbf{C}^{\mathrm{T}} \\ \mathbf{C} & \mathbf{B}^{\mathrm{e}} \end{pmatrix}$ with transition amplitudes $\mathbf{X} = \begin{pmatrix} \mathbf{X}^{\mathrm{p}} \\ \mathbf{X}^{\mathrm{e}} \end{pmatrix}$ and $\mathbf{Y} = \begin{pmatrix} \mathbf{Y}^{\mathrm{p}} \\ \mathbf{Y}^{\mathrm{e}} \end{pmatrix}$. Using the same technique as used to formulate the conventional electronic TDDFT Lagrangian,[60] we can express the NEO-TDDFT Lagrangian as

$$L[\mathbf{X},\mathbf{Y},\omega,\mathbf{C}^{\mathrm{e}},\mathbf{Z}^{\mathrm{e}},\mathbf{W}^{\mathrm{e}},\mathbf{C}^{\mathrm{p}},\mathbf{Z}^{\mathrm{p}},\mathbf{W}^{\mathrm{p}}] = G[\mathbf{X},\mathbf{Y},\omega] + \sum_{ia\sigma} Z^{\mathrm{e}}_{ia\sigma} F^{\mathrm{e}}_{ia\sigma} - \sum_{pq\sigma,p\le q} W^{\mathrm{e}}_{pq\sigma}\left(S^{\mathrm{e}}_{pq\sigma} - \delta_{pq}\right) + \sum_{i'a'} Z^{\mathrm{p}}_{i'a'} F^{\mathrm{p}}_{i'a'} - \sum_{p'q',p'\le q'} W^{\mathrm{p}}_{p'q'}\left(S^{\mathrm{p}}_{p'q'} - \delta_{p'q'}\right) \tag{7}$$

The first term $G[\mathbf{X},\mathbf{Y},\omega]$ is the Lagrangian for the NEO-TDDFT equations without the extra Lagrangian multipliers $\mathbf{Z}^{\mathrm{e}}$ ($\mathbf{Z}^{\mathrm{p}}$) and $\mathbf{W}^{\mathrm{e}}$ ($\mathbf{W}^{\mathrm{p}}$) for constraining the electronic (protonic) orbitals to be NEO-DFT solutions and to be orthonormal. Its specific form is

$$G[\mathbf{X},\mathbf{Y},\omega] = \frac{1}{2}\Big[\langle \mathbf{X}-\mathbf{Y}|\mathbf{A}-\mathbf{B}|\mathbf{X}-\mathbf{Y}\rangle - \omega\big(\langle \mathbf{X}-\mathbf{Y}|\mathbf{X}+\mathbf{Y}\rangle - 1\big) + \langle \mathbf{X}+\mathbf{Y}|\mathbf{A}+\mathbf{B}|\mathbf{X}+\mathbf{Y}\rangle - \omega\big(\langle \mathbf{X}+\mathbf{Y}|\mathbf{X}-\mathbf{Y}\rangle - 1\big)\Big] \tag{8}$$

which resembles the conventional TDDFT Lagrangian but with different definitions of the constituents. Its stationary points give rise to the NEO-TDDFT excitation energies $\omega$ and transition amplitudes $\mathbf{X}$ and $\mathbf{Y}$.

The matrices $\mathbf{A}-\mathbf{B}$ and $\mathbf{A}+\mathbf{B}$ are composed of four sub-blocks distinguished by the particle type (i.e., ee, pp, ep, and pe) with elements defined as

$$\begin{aligned}
(A-B)_{ia\sigma,jb\sigma'} &= (F^{\mathrm{e}}_{ab\sigma}\delta_{ij} - F^{\mathrm{e}}_{ij\sigma}\delta_{ab})\delta_{\sigma\sigma'} + c_{\mathrm{x}}\delta_{\sigma\sigma'}\big[(ja\sigma|ib\sigma) - (ab\sigma|ij\sigma)\big] \\
(A-B)_{i'a',j'b'} &= (F^{\mathrm{p}}_{a'b'}\delta_{i'j'} - F^{\mathrm{p}}_{i'j'}\delta_{a'b'}) + (j'a'|i'b') - (a'b'|i'j') \\
(A-B)_{ia\sigma,j'b'} &= (A-B)_{i'a',jb\sigma} = 0 \\
(A+B)_{ia\sigma,jb\sigma'} &= (F^{\mathrm{e}}_{ab\sigma}\delta_{ij} - F^{\mathrm{e}}_{ij\sigma}\delta_{ab})\delta_{\sigma\sigma'} + 2(ia\sigma|jb\sigma') - c_{\mathrm{x}}\delta_{\sigma\sigma'}\big[(ja\sigma|ib\sigma) + (ab\sigma|ij\sigma)\big] \\
&\quad + 2f^{\mathrm{exc}}_{ia\sigma,jb\sigma'} + 2f^{\mathrm{epc,ee}}_{ia\sigma,jb\sigma'} \\
(A+B)_{i'a',j'b'} &= (F^{\mathrm{p}}_{a'b'}\delta_{i'j'} - F^{\mathrm{p}}_{i'j'}\delta_{a'b'}) + 2(i'a'|j'b') - (j'a'|i'b') - (a'b'|i'j') + 2f^{\mathrm{epc,pp}}_{i'a',j'b'} \\
(A+B)_{ia\sigma,j'b'} &= -2(ia\sigma|j'b') + 2f^{\mathrm{epc,pe}}_{ia\sigma,j'b'} \\
(A+B)_{i'a',jb\sigma} &= -2(i'a'|jb\sigma) + 2f^{\mathrm{epc,ep}}_{i'a',jb\sigma}
\end{aligned} \tag{9}$$

Here $f^{\mathrm{exc}}_{pq\sigma,rs\sigma'}$ is a matrix element of the exc kernel in terms of the electronic orbitals $\psi^{\mathrm{e}}_{p}$, where the exc kernel is defined as $f^{\mathrm{exc}}_{\sigma\sigma'} = \dfrac{\delta^2 E^{\mathrm{exc}}}{\delta\rho^{\mathrm{e}}_{\sigma'}\delta\rho^{\mathrm{e}}_{\sigma}}$. The matrix elements of the epc kernels are defined analogously with respect to the electronic and protonic orbitals, where the epc kernels are defined as $f_{\sigma\sigma'}^{\mathrm{epc,ee}} = \frac{\delta^2 E^{\mathrm{epc}}}{\delta\rho_{\sigma'}^{\mathrm{e}}\delta\rho_{\sigma}^{\mathrm{e}}}$, $f^{\mathrm{epc,pp}} = \frac{\delta^2 E^{\mathrm{epc}}}{\delta\rho^{\mathrm{p}\,2}}$, and $f_{\sigma}^{\mathrm{epc,ep}} = f_{\sigma}^{\mathrm{epc,pe}} = \frac{\delta^2 E^{\mathrm{epc}}}{\delta\rho^{\mathrm{p}}\delta\rho_{\sigma}^{\mathrm{e}}}$. Introducing the constraints on the orbital coefficients in the NEO-TDDFT Lagrangian in Eq. (7) through the Lagrange multipliers $\mathbf{Z}^{\mathrm{e}}$, $\mathbf{Z}^{\mathrm{p}}$, $\mathbf{W}^{\mathrm{e}}$, and $\mathbf{W}^{\mathrm{p}}$ avoids solving the dependence of the orbital coefficients on the nuclear coordinates $3N$ times, where $N$ is the number of symmetry-unique atoms, and hence is more efficient. With these constraints on the coefficient matrices, $F_{pq\sigma}^{\mathrm{e}} = \delta_{pq}\varepsilon_{p\sigma}^{\mathrm{e}}$ and $F_{p'q'}^{\mathrm{p}} = \delta_{p'q'}\varepsilon_{p'}^{\mathrm{p}}$ for the various indices in Eq. (9) .

### 2.3 Solving the NEO-TDDFT Lagrangian

To solve for each variable of the Lagrangian in Eq. (7), we need to find the stationary points of the Lagrangian with respect to each variable. The orbital coefficients $\mathbf{C}^{\mathrm{e}}$ and $\mathbf{C}^{\mathrm{p}}$ are obtained by solving the NEO-DFT equations, whereas $\omega$, $\mathbf{X}$, and $\mathbf{Y}$ are obtained by solving the NEO-TDDFT equations. By finding the stationary points of the Lagrangian with respect to the orbital coefficients, the relations between $\mathbf{Z}^{\mathrm{e}}$ ($\mathbf{Z}^{\mathrm{p}}$) and $\mathbf{W}^{\mathrm{e}}$ ($\mathbf{W}^{\mathrm{p}}$) and a set of coupled Z vector equations are obtained. The latter can be used to solve for the Lagrange multipliers $\mathbf{Z}^{\mathrm{e}}$ and $\mathbf{Z}^{\mathrm{p}}$, and then the former can be used to obtain $\mathbf{W}^{\mathrm{e}}$ and $\mathbf{W}^{\mathrm{p}}$. The detailed derivation of the derivatives of the Lagrangian with respect to the orbital coefficients and the Z vector equations are provided in Section S1 of the SI. The Z vector equations are given as

$$\begin{aligned}
&\sum_{jb\sigma'}\left(A+B\right)_{ia\sigma,jb\sigma'} Z_{jb\sigma'}^{\mathrm{e}} + \sum_{i'a'}\left(A+B\right)_{ia\sigma,i'a'} Z_{i'a'}^{\mathrm{p}} = -R_{ia\sigma}^{\mathrm{e}} \\
&\sum_{j'b'}\left(A+B\right)_{i'a',j'b'} Z_{j'b'}^{\mathrm{p}} + \sum_{ia\sigma}\left(A+B\right)_{i'a',ia\sigma} Z_{ia\sigma}^{\mathrm{e}} = -R_{i'a'}^{\mathrm{p}}
\end{aligned} \tag{10}$$

The matrix elements of $\mathbf{R}^{\mathrm{e}}$ and $\mathbf{R}^{\mathrm{p}}$ are defined as

$$
\begin{aligned}
R_{ia\sigma}^{\mathrm{e}} &= H_{ia\sigma}^{\mathrm{e}+}\left[\mathbf{T}^{\mathrm{e}}\right]+H_{ia\sigma}^{\mathrm{ep}}\left[\mathbf{T}^{\mathrm{p}}\right] \\
&+\sum_{b}\left\{\left(X^{\mathrm{e}}+Y^{\mathrm{e}}\right)_{ib\sigma}\left(H_{ab\sigma}^{\mathrm{e}+}\left[\mathbf{X}^{\mathrm{e}}+\mathbf{Y}^{\mathrm{e}}\right]+H_{ab\sigma}^{\mathrm{ep}}\left[\mathbf{X}^{\mathrm{p}}+\mathbf{Y}^{\mathrm{p}}\right]\right)+\left(X^{\mathrm{e}}-Y^{\mathrm{e}}\right)_{ib\sigma}H_{ab\sigma}^{\mathrm{e}-}\left[\mathbf{X}^{\mathrm{e}}-\mathbf{Y}^{\mathrm{e}}\right]\right\} \\
&-\sum_{j}\left\{\left(X^{\mathrm{e}}+Y^{\mathrm{e}}\right)_{ja\sigma}\left(H_{ji\sigma}^{\mathrm{e}+}\left[\mathbf{X}^{\mathrm{e}}+\mathbf{Y}^{\mathrm{e}}\right]+H_{ji\sigma}^{\mathrm{ep}}\left[\mathbf{X}^{\mathrm{p}}+\mathbf{Y}^{\mathrm{p}}\right]\right)+\left(X^{\mathrm{e}}-Y^{\mathrm{e}}\right)_{ja\sigma}H_{ji\sigma}^{\mathrm{e}-}\left[\mathbf{X}^{\mathrm{e}}-\mathbf{Y}^{\mathrm{e}}\right]\right\} \\
&+\sum_{kc\sigma'}\left(X^{\mathrm{e}}+Y^{\mathrm{e}}\right)_{kc\sigma'}\left[\sum_{jb\sigma''}2\left(g_{kc\sigma',jb\sigma'',ia\sigma}^{\mathrm{exc}}+g_{kc\sigma',jb\sigma'',ia\sigma}^{\mathrm{epc,eee}}\right)\left(X^{\mathrm{e}}+Y^{\mathrm{e}}\right)_{jb\sigma''}+\sum_{i'a'}2g_{kc\sigma',i'a',ia\sigma}^{\mathrm{epc,epe}}\left(X^{\mathrm{p}}+Y^{\mathrm{p}}\right)_{i'a'}\right] \\
&+\sum_{i'a'}\left(X^{\mathrm{p}}+Y^{\mathrm{p}}\right)_{i'a'}\left[\sum_{jb\sigma'}2g_{i'a',jb\sigma',ia\sigma}^{\mathrm{epc,eep}}\left(X^{\mathrm{e}}+Y^{\mathrm{e}}\right)_{jb\sigma'}+\sum_{j'b'}2g_{i'a',j'b',ia\sigma}^{\mathrm{epc,epp}}\left(X^{\mathrm{p}}+Y^{\mathrm{p}}\right)_{j'b'}\right]
\end{aligned}
$$

$$
\begin{aligned}
R_{i'a'}^{\mathrm{p}} &= H_{i'a'}^{\mathrm{p}+}\left[\mathbf{T}^{\mathrm{p}}\right]+H_{i'a'}^{\mathrm{pe}}\left[\mathbf{T}^{\mathrm{e}}\right] \\
&+\sum_{b'}\left\{\left(X^{\mathrm{p}}+Y^{\mathrm{p}}\right)_{i'b'}\left(H_{a'b'}^{\mathrm{p}+}\left[\mathbf{X}^{\mathrm{p}}+\mathbf{Y}^{\mathrm{p}}\right]+H_{a'b'}^{\mathrm{pe}}\left[\mathbf{X}^{\mathrm{e}}+\mathbf{Y}^{\mathrm{e}}\right]\right)+\left(X^{\mathrm{p}}-Y^{\mathrm{p}}\right)_{i'b'}H_{a'b'}^{\mathrm{p}-}\left[\mathbf{X}^{\mathrm{p}}-\mathbf{Y}^{\mathrm{p}}\right]\right\} \\
&-\sum_{j'}\left\{\left(X^{\mathrm{p}}+Y^{\mathrm{p}}\right)_{j'a'}\left(H_{j'i'}^{\mathrm{p}+}\left[\mathbf{X}^{\mathrm{p}}+\mathbf{Y}^{\mathrm{p}}\right]+H_{j'i'}^{\mathrm{pe}}\left[\mathbf{X}^{\mathrm{e}}+\mathbf{Y}^{\mathrm{e}}\right]\right)+\left(X^{\mathrm{p}}-Y^{\mathrm{p}}\right)_{j'a'}H_{j'i'}^{\mathrm{p}-}\left[\mathbf{X}^{\mathrm{p}}-\mathbf{Y}^{\mathrm{p}}\right]\right\} \\
&+\sum_{k'c'}\left(X^{\mathrm{p}}+Y^{\mathrm{p}}\right)_{k'c'}\left[\sum_{j'b'}2g_{k'c',j'b',i'a'}^{\mathrm{epc,ppp}}\left(X^{\mathrm{p}}+Y^{\mathrm{p}}\right)_{j'b'}+\sum_{ia\sigma}2g_{k'c',ia\sigma,i'a'}^{\mathrm{epc,pep}}\left(X^{\mathrm{e}}+Y^{\mathrm{e}}\right)_{ia\sigma}\right] \\
&+\sum_{ia\sigma}\left(X^{\mathrm{e}}+Y^{\mathrm{e}}\right)_{ia\sigma}\left[\sum_{j'b'}2g_{ia\sigma,j'b',i'a'}^{\mathrm{ep,ppe}}\left(X^{\mathrm{p}}+Y^{\mathrm{p}}\right)_{j'b'}+\sum_{jb\sigma'}2g_{ia\sigma,jb\sigma',i'a'}^{\mathrm{epc,pee}}\left(X^{\mathrm{e}}+Y^{\mathrm{e}}\right)_{jb\sigma'}\right]
\end{aligned} \tag{11}
$$

Here, the unrelaxed difference density $\mathbf{T}^{\mathrm{e}}$ has matrix elements

$$
\begin{aligned}
T_{ij\sigma}^{\mathrm{e}} &= -\frac{1}{2}\sum_{a}\left[\left(X^{\mathrm{e}}-Y^{\mathrm{e}}\right)_{ia\sigma}\left(X^{\mathrm{e}}-Y^{\mathrm{e}}\right)_{ja\sigma}+\left(X^{\mathrm{e}}+Y^{\mathrm{e}}\right)_{ia\sigma}\left(X^{\mathrm{e}}+Y^{\mathrm{e}}\right)_{ja\sigma}\right] \\
T_{ab\sigma}^{\mathrm{e}} &= \frac{1}{2}\sum_{i}\left[\left(X^{\mathrm{e}}-Y^{\mathrm{e}}\right)_{ia\sigma}\left(X^{\mathrm{e}}-Y^{\mathrm{e}}\right)_{ib\sigma}+\left(X^{\mathrm{e}}+Y^{\mathrm{e}}\right)_{ia\sigma}\left(X^{\mathrm{e}}+Y^{\mathrm{e}}\right)_{ib\sigma}\right] \\
T_{ia\sigma}^{\mathrm{e}} &= T_{ai\sigma}^{\mathrm{e}} = 0
\end{aligned} \tag{12}
$$

The corresponding protonic difference density $\mathbf{T}^{\mathrm{p}}$ is defined analogously. Here $g_{pq\sigma,rs\sigma',tu\sigma''}^{\mathrm{exc}}$ is a matrix element of the third-order functional derivative of the exc energy, which is defined as $g_{\sigma'\sigma''\sigma}^{\mathrm{exc}} = \dfrac{\delta^3 E^{\mathrm{exc}}}{\delta\rho_{\sigma'}^{\mathrm{e}}\delta\rho_{\sigma''}^{\mathrm{e}}\delta\rho_{\sigma}^{\mathrm{e}}}$, with respect to the electronic orbitals. The third-order functional derivatives of the epc energy are defined analogously and can be distinguished by the labels, e.g., $g_{\sigma}^{\mathrm{epc,epp}} = \dfrac{\delta^3 E^{\mathrm{epc}}}{\delta \rho^{\mathrm{p}\,2} \delta \rho_{\sigma}^{\mathrm{e}}}$, and the indices of the matrix elements identify the electronic and protonic orbitals. The linear transforms used in Eq. (11) are defined in terms of a general matrix $\mathbf{V}^{\mathrm{e}}$ or $\mathbf{V}^{\mathrm{p}}$ as

$$
\begin{aligned}
H_{pq\sigma}^{\mathrm{e+}}\left[\mathbf{V}^{\mathrm{e}}\right] &= \sum_{rs\sigma'} \left\{ 2\left(pq\sigma \,|\, rs\sigma'\right) - c_{\mathrm{x}} \delta_{\sigma\sigma'} \left[ \left(ps\sigma \,|\, rq\sigma\right) + \left(pr\sigma \,|\, sq\sigma\right) \right] + 2 f_{pq\sigma,rs\sigma'}^{\mathrm{exc}} + 2 f_{pq\sigma,rs\sigma'}^{\mathrm{epc,ee}} \right\} V_{rs\sigma'}^{\mathrm{e}} \\
H_{pq\sigma}^{\mathrm{e-}}\left[\mathbf{V}^{\mathrm{e}}\right] &= \sum_{rs\sigma'} c_{\mathrm{x}} \delta_{\sigma\sigma'} \left[ \left(ps\sigma \,|\, rq\sigma\right) - \left(pr\sigma \,|\, sq\sigma\right) \right] V_{rs\sigma'}^{\mathrm{e}} \\
H_{pq\sigma}^{\mathrm{ep}}\left[\mathbf{V}^{\mathrm{p}}\right] &= \sum_{p'q'} \left(A+B\right)_{pq\sigma,p'q'} V_{p'q'}^{\mathrm{p}} \\
H_{p'q'}^{\mathrm{p+}}\left[\mathbf{V}^{\mathrm{p}}\right] &= \sum_{r's'} \left[ 2\left(p'q' \,|\, r's'\right) - \left(p's' \,|\, r'q'\right) - \left(p'r' \,|\, s'q'\right) + 2 f_{p'q',r's'}^{\mathrm{epc,pp}} \right] V_{r's'}^{\mathrm{p}} \\
H_{p'q'}^{\mathrm{p-}}\left[\mathbf{V}^{\mathrm{p}}\right] &= \sum_{r's'} \left[ \left(p's' \,|\, r'q'\right) - \left(p'r' \,|\, s'q'\right) \right] V_{r's'}^{\mathrm{p}} \\
H_{p'q'}^{\mathrm{pe}}\left[\mathbf{V}^{\mathrm{e}}\right] &= \sum_{pq\sigma} \left(A+B\right)_{p'q',pq\sigma} V_{pq\sigma}^{\mathrm{e}}
\end{aligned}
\tag{13}
$$

After calculating $\mathbf{Z}^{\mathrm{e}}$ and $\mathbf{Z}^{\mathrm{p}}$, we can define the relaxed one-particle difference density matrix for each type of particle as $\mathbf{P}^{\mathrm{e}} = \mathbf{T}^{\mathrm{e}} + \mathbf{Z}^{\mathrm{e}}$ and $\mathbf{P}^{\mathrm{p}} = \mathbf{T}^{\mathrm{p}} + \mathbf{Z}^{\mathrm{p}}$. Then $\mathbf{W}^{\mathrm{e}}$ and $\mathbf{W}^{\mathrm{p}}$ can be obtained from the following equations:

$$
\begin{aligned}
W^{\mathrm{e}}_{ij\sigma} = \frac{1}{\left(\delta_{ij}+1\right)} \Bigg\{ & H^{\mathrm{e}+}_{ij\sigma}\left[\mathbf{P}^{\mathrm{e}}\right] + H^{\mathrm{ep}}_{ij\sigma}\left[\mathbf{P}^{\mathrm{p}}\right] \\
& + \sum_{a} \Big\{ \omega \Big[ \left(X^{\mathrm{e}}-Y^{\mathrm{e}}\right)_{ia\sigma} \left(X^{\mathrm{e}}+Y^{\mathrm{e}}\right)_{ja\sigma} + \left(X^{\mathrm{e}}+Y^{\mathrm{e}}\right)_{ia\sigma} \left(X^{\mathrm{e}}-Y^{\mathrm{e}}\right)_{ja\sigma} \Big] \\
& - \varepsilon^{\mathrm{e}}_{a\sigma} \Big[ \left(X^{\mathrm{e}}-Y^{\mathrm{e}}\right)_{ia\sigma} \left(X^{\mathrm{e}}-Y^{\mathrm{e}}\right)_{ja\sigma} + \left(X^{\mathrm{e}}+Y^{\mathrm{e}}\right)_{ia\sigma} \left(X^{\mathrm{e}}+Y^{\mathrm{e}}\right)_{ja\sigma} \Big] \Big\} \\
& + \sum_{ka\sigma'} \left(X^{\mathrm{e}}+Y^{\mathrm{e}}\right)_{ka\sigma'} \Bigg[ \sum_{lb\sigma''} 2\left( g^{\mathrm{exc}}_{ka\sigma',lb\sigma'',ij\sigma} + g^{\mathrm{epc,eee}}_{ka\sigma',lb\sigma'',ij\sigma} \right) \left(X^{\mathrm{e}}+Y^{\mathrm{e}}\right)_{lb\sigma''} + \sum_{i'a'} 2 g^{\mathrm{epc,epe}}_{ka\sigma',i'a',ij\sigma} \left(X^{\mathrm{p}}+Y^{\mathrm{p}}\right)_{i'a'} \Bigg] \\
& + \sum_{i'a'} \left(X^{\mathrm{p}}+Y^{\mathrm{p}}\right)_{i'a'} \Bigg[ \sum_{ka\sigma'} 2 g^{\mathrm{epc,eep}}_{i'a',ka\sigma',ij\sigma} \left(X^{\mathrm{e}}+Y^{\mathrm{e}}\right)_{ka\sigma'} + \sum_{j'b'} 2 g^{\mathrm{epc,epp}}_{i'a',j'b',ij\sigma} \left(X^{\mathrm{p}}+Y^{\mathrm{p}}\right)_{j'b'} \Bigg] \Bigg\}
\end{aligned}
$$

$$
\begin{aligned}
W^{\mathrm{e}}_{ab\sigma} = \frac{1}{\left(\delta_{ab}+1\right)} \sum_{i} \Big\{ & \omega \Big[ \left(X^{\mathrm{e}}-Y^{\mathrm{e}}\right)_{ia\sigma} \left(X^{\mathrm{e}}+Y^{\mathrm{e}}\right)_{ib\sigma} + \left(X^{\mathrm{e}}+Y^{\mathrm{e}}\right)_{ia\sigma} \left(X^{\mathrm{e}}-Y^{\mathrm{e}}\right)_{ib\sigma} \Big] \\
& + \varepsilon^{\mathrm{e}}_{i\sigma} \Big[ \left(X^{\mathrm{e}}-Y^{\mathrm{e}}\right)_{ia\sigma} \left(X^{\mathrm{e}}-Y^{\mathrm{e}}\right)_{ib\sigma} + \left(X^{\mathrm{e}}+Y^{\mathrm{e}}\right)_{ia\sigma} \left(X^{\mathrm{e}}+Y^{\mathrm{e}}\right)_{ib\sigma} \Big] \Big\}
\end{aligned}
$$

$$
W^{\mathrm{e}}_{ia\sigma} = \sum_{j} \Big\{ \left(X^{\mathrm{e}}+Y^{\mathrm{e}}\right)_{ja\sigma} \left( H^{\mathrm{e}+}_{ji\sigma}\left[\mathbf{X}^{\mathrm{e}}+\mathbf{Y}^{\mathrm{e}}\right] + H^{\mathrm{ep}}_{ji\sigma}\left[\mathbf{X}^{\mathrm{p}}+\mathbf{Y}^{\mathrm{p}}\right] \right) + \left(X^{\mathrm{e}}-Y^{\mathrm{e}}\right)_{ja\sigma} \left( H^{\mathrm{e}-}_{ji\sigma}\left[\mathbf{X}^{\mathrm{e}}-\mathbf{Y}^{\mathrm{e}}\right] \right) \Big\} + Z^{\mathrm{e}}_{ia\sigma} \varepsilon^{\mathrm{e}}_{i\sigma}
$$

$$
\begin{aligned}
W^{\mathrm{p}}_{i'j'} = \frac{1}{\left(\delta_{i'j'}+1\right)} \Bigg\{ & H^{\mathrm{p}+}_{i'j'}\left[\mathbf{P}^{\mathrm{p}}\right] + H^{\mathrm{pe}}_{i'j'}\left[\mathbf{P}^{\mathrm{e}}\right] \\
& + \sum_{a'} \Big\{ \omega \Big[ \left(X^{\mathrm{p}}-Y^{\mathrm{p}}\right)_{i'a'} \left(X^{\mathrm{p}}+Y^{\mathrm{p}}\right)_{j'a'} + \left(X^{\mathrm{p}}+Y^{\mathrm{p}}\right)_{i'a'} \left(X^{\mathrm{p}}-Y^{\mathrm{p}}\right)_{j'a'} \Big] \\
& - \varepsilon^{\mathrm{p}}_{a'} \Big[ \left(X^{\mathrm{p}}-Y^{\mathrm{p}}\right)_{i'a'} \left(X^{\mathrm{p}}-Y^{\mathrm{p}}\right)_{j'a'} + \left(X^{\mathrm{p}}+Y^{\mathrm{p}}\right)_{i'a'} \left(X^{\mathrm{p}}+Y^{\mathrm{p}}\right)_{j'a'} \Big] \Big\} \\
& + \sum_{k'a'} \left(X^{\mathrm{p}}+Y^{\mathrm{p}}\right)_{k'a'} \Bigg[ \sum_{l'b'} 2 g^{\mathrm{epc,ppp}}_{k'a',l'b',i'j'} \left(X^{\mathrm{p}}+Y^{\mathrm{p}}\right)_{l'b'} + \sum_{ia\sigma} 2 g^{\mathrm{epc,pep}}_{k'a',ia\sigma,i'j'} \left(X^{\mathrm{e}}+Y^{\mathrm{e}}\right)_{ia\sigma} \Bigg] \\
& + \sum_{ia\sigma} \left(X^{\mathrm{e}}+Y^{\mathrm{e}}\right)_{ia\sigma} \Bigg[ \sum_{k'a'} 2 g^{\mathrm{epc,ppe}}_{ia\sigma,k'a',i'j'} \left(X^{\mathrm{p}}+Y^{\mathrm{p}}\right)_{k'a'} + \sum_{jb\sigma'} 2 g^{\mathrm{epc,pee}}_{ia\sigma,jb\sigma',i'j'} \left(X^{\mathrm{e}}+Y^{\mathrm{e}}\right)_{jb\sigma'} \Bigg] \Bigg\}
\end{aligned}
$$

$$
\begin{aligned}
W^{\mathrm{p}}_{a'b'} = \frac{1}{\left(\delta_{a'b'}+1\right)} \sum_{i'} \Big\{ & \omega \Big[ \left(X^{\mathrm{p}}-Y^{\mathrm{p}}\right)_{i'a'} \left(X^{\mathrm{p}}+Y^{\mathrm{p}}\right)_{i'b'} + \left(X^{\mathrm{p}}+Y^{\mathrm{p}}\right)_{i'a'} \left(X^{\mathrm{p}}-Y^{\mathrm{p}}\right)_{i'b'} \Big] \\
& + \varepsilon^{\mathrm{p}}_{i'} \Big[ \left(X^{\mathrm{p}}-Y^{\mathrm{p}}\right)_{i'a'} \left(X^{\mathrm{p}}-Y^{\mathrm{p}}\right)_{i'b'} + \left(X^{\mathrm{p}}+Y^{\mathrm{p}}\right)_{i'a'} \left(X^{\mathrm{p}}+Y^{\mathrm{p}}\right)_{i'b'} \Big] \Big\}
\end{aligned} \tag{14}
$$

$$
W^{\mathrm{p}}_{i'a'} = \sum_{j'} \Big\{ \left(X^{\mathrm{p}}+Y^{\mathrm{p}}\right)_{j'a'} \left( H^{\mathrm{p}+}_{j'i'}\left[\mathbf{X}^{\mathrm{p}}+\mathbf{Y}^{\mathrm{p}}\right] + H^{\mathrm{pe}}_{j'i'}\left[\mathbf{X}^{\mathrm{e}}+\mathbf{Y}^{\mathrm{e}}\right] \right) + \left(X^{\mathrm{p}}-Y^{\mathrm{p}}\right)_{j'a'} \left( H^{\mathrm{p}-}_{j'i'}\left[\mathbf{X}^{\mathrm{p}}-\mathbf{Y}^{\mathrm{p}}\right] \right) \Big\} + Z^{\mathrm{p}}_{i'a'} \varepsilon^{\mathrm{p}}_{i'}
$$

## 2.4 NEO-TDDFT Gradient

With the full Lagrangian solved, we obtain the NEO-TDDFT gradient by taking the derivative of the Lagrangian with respect to a nuclear displacement $\xi$:

$$
\begin{aligned}
L^{\xi} &= \sum_{\mu\nu\sigma} h^{\mathrm{e}\xi}_{\mu\nu} P^{\mathrm{e}}_{\mu\nu\sigma} + \sum_{\mu'\nu'} h^{\mathrm{p}\xi}_{\mu'\nu'} P^{\mathrm{p}}_{\mu'\nu'} - \sum_{\mu\nu\sigma} S^{\mathrm{e}\xi}_{\mu\nu} W^{\mathrm{e}}_{\mu\nu\sigma} - \sum_{\mu'\nu'} S^{\mathrm{p}\xi}_{\mu'\nu'} W^{\mathrm{p}}_{\mu'\nu'} \\
&+ \sum_{\mu\nu\sigma\lambda\kappa\sigma'} (\mu\nu|\lambda\kappa)^{\xi} \Gamma^{\mathrm{ee}}_{\mu\nu\sigma,\lambda\kappa\sigma'} + \sum_{\mu'\nu'\lambda'\kappa'} (\mu'\nu'|\lambda'\kappa')^{\xi} \Gamma^{\mathrm{pp}}_{\mu'\nu',\lambda'\kappa'} - \sum_{\mu\nu\sigma\mu'\nu'} (\mu\nu|\mu'\nu')^{\xi} \Gamma^{\mathrm{ep}}_{\mu\nu\sigma,\mu'\nu'} \\
&+ \sum_{\mu\nu\sigma} v^{\mathrm{exc}(\xi)}_{\mu\nu\sigma} P^{\mathrm{e}}_{\mu\nu\sigma} + \sum_{\mu\nu\sigma} v^{\mathrm{epc,e}(\xi)}_{\mu\nu\sigma} P^{\mathrm{e}}_{\mu\nu\sigma} + \sum_{\mu'\nu'} v^{\mathrm{epc,p}(\xi)}_{\mu'\nu'} P^{\mathrm{p}}_{\mu'\nu'} \\
&+ \sum_{\mu\nu\sigma} (X^{\mathrm{e}}+Y^{\mathrm{e}})_{\mu\nu\sigma} \left[ \sum_{\lambda\kappa\sigma'} \left( f^{\mathrm{exc}(\xi)}_{\mu\nu\sigma,\lambda\kappa\sigma'} + f^{\mathrm{epc,ee}(\xi)}_{\mu\nu\sigma,\lambda\kappa\sigma'} \right) (X^{\mathrm{e}}+Y^{\mathrm{e}})_{\lambda\kappa\sigma'} + \sum_{\mu'\nu'} f^{\mathrm{epc,pe}(\xi)}_{\mu\nu\sigma,\mu'\nu'} (X^{\mathrm{p}}+Y^{\mathrm{p}})_{\mu'\nu'} \right] \\
&+ \sum_{\mu'\nu'} (X^{\mathrm{p}}+Y^{\mathrm{p}})_{\mu'\nu'} \left[ \sum_{\lambda'\kappa'} f^{\mathrm{epc,pp}(\xi)}_{\mu'\nu'\lambda'\kappa'} (X^{\mathrm{p}}+Y^{\mathrm{p}})_{\lambda'\kappa'} + \sum_{\mu\nu\sigma} f^{\mathrm{epc,ep}(\xi)}_{\mu'\nu',\mu\nu\sigma} (X^{\mathrm{e}}+Y^{\mathrm{e}})_{\mu\nu\sigma} \right]
\end{aligned}
\tag{15}
$$

Here $(\xi)$ indicates taking the derivative with respect to the nuclear displacement $\xi$ with the orbital coefficients fixed. All of the matrix elements in this subsection are given with respect to the electronic and protonic basis functions $\phi^{\mathrm{e}}_{\mu}$ and $\phi^{\mathrm{p}}_{\mu'}$. The effective two-particle difference density matrices are defined as

$$
\begin{aligned}
\Gamma^{\mathrm{ee}}_{\mu\nu\sigma,\lambda\kappa\sigma'} &= P^{\mathrm{e}}_{\mu\nu\sigma} D^{\mathrm{e}}_{\lambda\kappa\sigma'} + \left(X^{\mathrm{e}}+Y^{\mathrm{e}}\right)_{\mu\nu\sigma} \left(X^{\mathrm{e}}+Y^{\mathrm{e}}\right)_{\lambda\kappa\sigma'} - \frac{c_{\mathrm{x}}}{2} \delta_{\sigma\sigma'} \Big[ P^{\mathrm{e}}_{\mu\kappa\sigma} D^{\mathrm{e}}_{\lambda\nu\sigma'} + P^{\mathrm{e}}_{\mu\lambda\sigma} D^{\mathrm{e}}_{\kappa\nu\sigma'} + \left(X^{\mathrm{e}}+Y^{\mathrm{e}}\right)_{\mu\kappa\sigma} \left(X^{\mathrm{e}}+Y^{\mathrm{e}}\right)_{\lambda\nu\sigma'} \\
&+ \left(X^{\mathrm{e}}+Y^{\mathrm{e}}\right)_{\mu\lambda\sigma} \left(X^{\mathrm{e}}+Y^{\mathrm{e}}\right)_{\kappa\nu\sigma'} - \left(X^{\mathrm{e}}-Y^{\mathrm{e}}\right)_{\mu\kappa\sigma} \left(X^{\mathrm{e}}-Y^{\mathrm{e}}\right)_{\lambda\nu\sigma'} + \left(X^{\mathrm{e}}-Y^{\mathrm{e}}\right)_{\mu\lambda\sigma} \left(X^{\mathrm{e}}+Y^{\mathrm{e}}\right)_{\kappa\nu\sigma'} \Big] \\
\Gamma^{\mathrm{pp}}_{\mu'\nu',\lambda'\kappa'} &= P^{\mathrm{p}}_{\mu'\nu'} D^{\mathrm{p}}_{\lambda'\kappa'} + \left(X^{\mathrm{p}}+Y^{\mathrm{p}}\right)_{\mu'\nu'} \left(X^{\mathrm{p}}+Y^{\mathrm{p}}\right)_{\lambda'\kappa'} - \frac{1}{2} \Big[ P^{\mathrm{p}}_{\mu'\kappa'} D^{\mathrm{p}}_{\lambda'\nu'} + P^{\mathrm{p}}_{\mu'\lambda'} D^{\mathrm{p}}_{\kappa'\nu'} + \left(X^{\mathrm{p}}+Y^{\mathrm{p}}\right)_{\mu'\kappa'} \left(X^{\mathrm{p}}+Y^{\mathrm{p}}\right)_{\lambda'\nu'} \\
&+ \left(X^{\mathrm{p}}+Y^{\mathrm{p}}\right)_{\mu'\lambda'} \left(X^{\mathrm{p}}+Y^{\mathrm{p}}\right)_{\kappa'\nu'} - \left(X^{\mathrm{p}}-Y^{\mathrm{p}}\right)_{\mu'\kappa'} \left(X^{\mathrm{p}}-Y^{\mathrm{p}}\right)_{\lambda'\nu'} + \left(X^{\mathrm{p}}-Y^{\mathrm{p}}\right)_{\mu'\lambda'} \left(X^{\mathrm{p}}-Y^{\mathrm{p}}\right)_{\kappa'\nu'} \Big] \\
\Gamma^{\mathrm{ep}}_{\mu\nu\sigma,\mu'\nu'} &= P^{\mathrm{e}}_{\mu\nu\sigma} D^{\mathrm{p}}_{\mu'\nu'} + P^{\mathrm{p}}_{\mu'\nu'} D^{\mathrm{e}}_{\mu\nu\sigma} + 2\left(X^{\mathrm{e}}+Y^{\mathrm{e}}\right)_{\mu\nu\sigma} \left(X^{\mathrm{p}}+Y^{\mathrm{p}}\right)_{\mu'\nu'}
\end{aligned}
\tag{16}
$$

To find the gradient of a specific NEO-TDDFT excited state, we need to add the excitation energy gradient $L^{\xi}$ to the NEO-DFT ground state gradient. The equation for the latter is

$$
\begin{aligned}
E^{\xi} &= E^{\mathrm{NN}\xi} + \sum_{\mu\nu\sigma} h^{\mathrm{e}\xi}_{\mu\nu} D^{\mathrm{e}}_{\mu\nu\sigma} + \sum_{\mu'\nu'} h^{\mathrm{p}\xi}_{\mu'\nu'} D^{\mathrm{p}}_{\mu'\nu'} - \sum_{\mu\nu\sigma} S^{\mathrm{e}\xi}_{\mu\nu} M^{\mathrm{e}}_{\mu\nu\sigma} - \sum_{\mu'\nu'} S^{\mathrm{p}\xi}_{\mu'\nu'} M^{\mathrm{p}}_{\mu'\nu'} \\
&+ \sum_{\mu\nu\sigma\lambda\kappa\sigma'} (\mu\nu|\lambda\kappa)^{\xi} G^{\mathrm{ee}}_{\mu\nu\sigma,\lambda\kappa\sigma'} + \sum_{\mu'\nu'\lambda'\kappa'} (\mu'\nu'|\lambda'\kappa')^{\xi} G^{\mathrm{pp}}_{\mu'\nu',\lambda'\kappa'} \\
&- \sum_{\mu\nu\sigma\mu'\nu'} (\mu\nu|\mu'\nu')^{\xi} G^{\mathrm{ep}}_{\mu\nu\sigma,\mu'\nu'} + E^{\mathrm{exc}(\xi)} + E^{\mathrm{epc}(\xi)}
\end{aligned} \tag{17}
$$

where $E^{\mathrm{NN}(\xi)}$ is the gradient of the Coulomb repulsion between the classical nuclei. $\mathbf{M}^{\mathrm{e}}$ and $\mathbf{M}^{\mathrm{p}}$ are the electronic and protonic energy-weighted density matrices defined as

$$
\begin{aligned}
M^{\mathrm{e}}_{\mu\nu\sigma} &= \sum_{i} \varepsilon^{\mathrm{e}}_{i\sigma} C^{\mathrm{e}}_{\mu i\sigma} C^{\mathrm{e}}_{\nu i\sigma} \\
M^{\mathrm{p}}_{\mu'\nu'} &= \sum_{i'} \varepsilon^{\mathrm{p}}_{i'} C^{\mathrm{p}}_{\mu' i'} C^{\mathrm{p}}_{\nu' i'}
\end{aligned} \tag{18}
$$

The effective ground state two-particle density matrices $\mathbf{G}^{\mathrm{ee}}$, $\mathbf{G}^{\mathrm{pp}}$, and $\mathbf{G}^{\mathrm{ep}}$ are defined as

$$
\begin{aligned}
G^{\mathrm{ee}}_{\mu\nu\sigma,\lambda\kappa\sigma'} &= \frac{1}{2}\left( D^{\mathrm{e}}_{\mu\nu\sigma} D^{\mathrm{e}}_{\lambda\kappa\sigma'} - c_{\mathrm{x}} \delta_{\sigma\sigma'} D^{\mathrm{e}}_{\mu\kappa\sigma} D^{\mathrm{e}}_{\lambda\nu\sigma} \right) \\
G^{\mathrm{pp}}_{\mu'\nu',\lambda'\kappa'} &= \frac{1}{2}\left( D^{\mathrm{p}}_{\mu'\nu'} D^{\mathrm{p}}_{\lambda'\kappa'} - D^{\mathrm{p}}_{\mu'\kappa'} D^{\mathrm{p}}_{\lambda'\nu'} \right) \\
G^{\mathrm{ep}}_{\mu\nu\sigma,\mu'\nu'} &= D^{\mathrm{e}}_{\mu\nu\sigma} D^{\mathrm{p}}_{\mu'\nu'}
\end{aligned} \tag{19}
$$

### 2.5 NEO-TDA Gradient

The TDA approximation has been applied to NEO-TDDFT previously[33] with the working equation $\mathbf{AX} = \omega\mathbf{X}$, where $\mathbf{A} = \begin{pmatrix} \mathbf{A}^{\mathrm{p}} & \mathbf{C}^{\mathrm{T}} \\ \mathbf{C} & \mathbf{A}^{\mathrm{e}} \end{pmatrix}$ and $\mathbf{X} = \begin{pmatrix} \mathbf{X}^{\mathrm{p}} \\ \mathbf{X}^{\mathrm{e}} \end{pmatrix}$. The Lagrangian approach used above to derive the NEO-TDDFT gradient can be used to derive the NEO-TDA gradient. The NEO-TDA Lagrangian is

$$
\begin{aligned}
&L[\mathbf{X},\omega,\mathbf{C}^{\mathrm{e}},\mathbf{Z}^{\mathrm{e}},\mathbf{W}^{\mathrm{e}},\mathbf{C}^{\mathrm{p}},\mathbf{Z}^{\mathrm{p}},\mathbf{W}^{\mathrm{p}}] \\
&= G[\mathbf{X},\omega] + \sum_{ia\sigma} Z^{\mathrm{e}}_{ia\sigma} F^{\mathrm{e}}_{ia\sigma} - \sum_{pq\sigma,p\le q} W^{\mathrm{e}}_{pq\sigma}\left( S^{\mathrm{e}}_{pq\sigma} - \delta_{pq} \right) + \sum_{i'a'} Z^{\mathrm{p}}_{i'a'} F^{\mathrm{p}}_{i'a'} - \sum_{p'q',p'\le q'} W^{\mathrm{p}}_{p'q'}\left( S^{\mathrm{p}}_{p'q'} - \delta_{p'q'} \right)
\end{aligned} \tag{20}
$$

where $G[\mathbf{X},\omega] = \langle \mathbf{X} | \mathbf{A} | \mathbf{X} \rangle - \omega(\langle \mathbf{X} | \mathbf{X} \rangle - 1)$. This Lagrangian does not depend on the de-excitation amplitudes. Similar procedures can be used to solve for each variable in the Lagrangian. Finding the stationary point with respect to the orbital coefficients gives rise to a set of Z vector equations, analogous to Eq.(10), but with different definitions for $\mathbf{R}^e$ and $\mathbf{R}^p$. The final analytical gradient for NEO-TDA can be also be formulated in a similar form as given in Eq. (15), with different expressions for the effective two-particle difference density matrices and no contributions from the de-excitation amplitudes $\mathbf{Y}^e$ and $\mathbf{Y}^p$. A complete derivation for the NEO-TDA gradient is presented in Section S2 of the SI.

## 3. Results and Discussion

We implemented the NEO-TDDFT and NEO-TDA analytical gradients in a developer version of Q-Chem 5.3.[61] These implementations have been validated by comparing analytical gradients to numerical gradients, as given in Section S3 of the SI. The NEO-TDDFT and NEO-TDA gradients were connected to the various geometry search algorithms in Q-Chem 5.3[61] to conduct geometry optimizations. The geometry optimizations were performed by minimizing the energy with respect to all classical nuclei as well as the basis function centers associated with the quantum protons, where the electronic and protonic basis functions for each quantum proton are assumed to be centered at the same position. The NEO-TDDFT and NEO-TDA solutions were solved with the Davidson algorithm,[62-63] and the Z vector equations were solved with the block conjugate gradient algorithm.[64]

### 3.1 0-0 Adiabatic Excitation Energies for Set of Nine Molecules

To calculate the 0-0 adiabatic excitation energies for the set of nine molecules, we first conducted geometry optimizations on the ground and select excited state surfaces with the NEO-TDDFT and NEO-TDA methods. The NEO-DFT ground and NEO-TDDFT excited state optimized geometries are depicted in Figure 1b and Figure 3, respectively. The enery difference between the optimized excited state and ground state geometries provides the adiabatic excitation energy. To calculate the 0-0 adiabatic excitation energy, we need to compute the ZPE difference between the excited state and the ground state optimized geometries. Because all protons are treated quantum mechanically, the ZPE associated with the quantum protons is included inherently in the NEO-TDDFT or NEO-TDA energy. The ZPE contributions from the classical nuclei were obtained by computing semi-numerical NEO Hessians on the NEO-TDDFT and NEO-TDA excited state surfaces and fully analytical NEO-DFT Hessians on the ground state surfaces. The equations for the analytical NEO-HF Hessian were published in Ref [32], and the equations for the analytical NEO-DFT Hessian are presented in Section S4 of the SI. The stationary points located on the NEO ground and excited state potential energy surfaces were shown to be minima by confirming that all eigenvalues of the Hessian were positive. Note that this procedure invokes the harmonic approximation for the frequencies associated with the classical nuclei, as well as the Born-Oppenheimer separation between the quantum and classical nuclei.

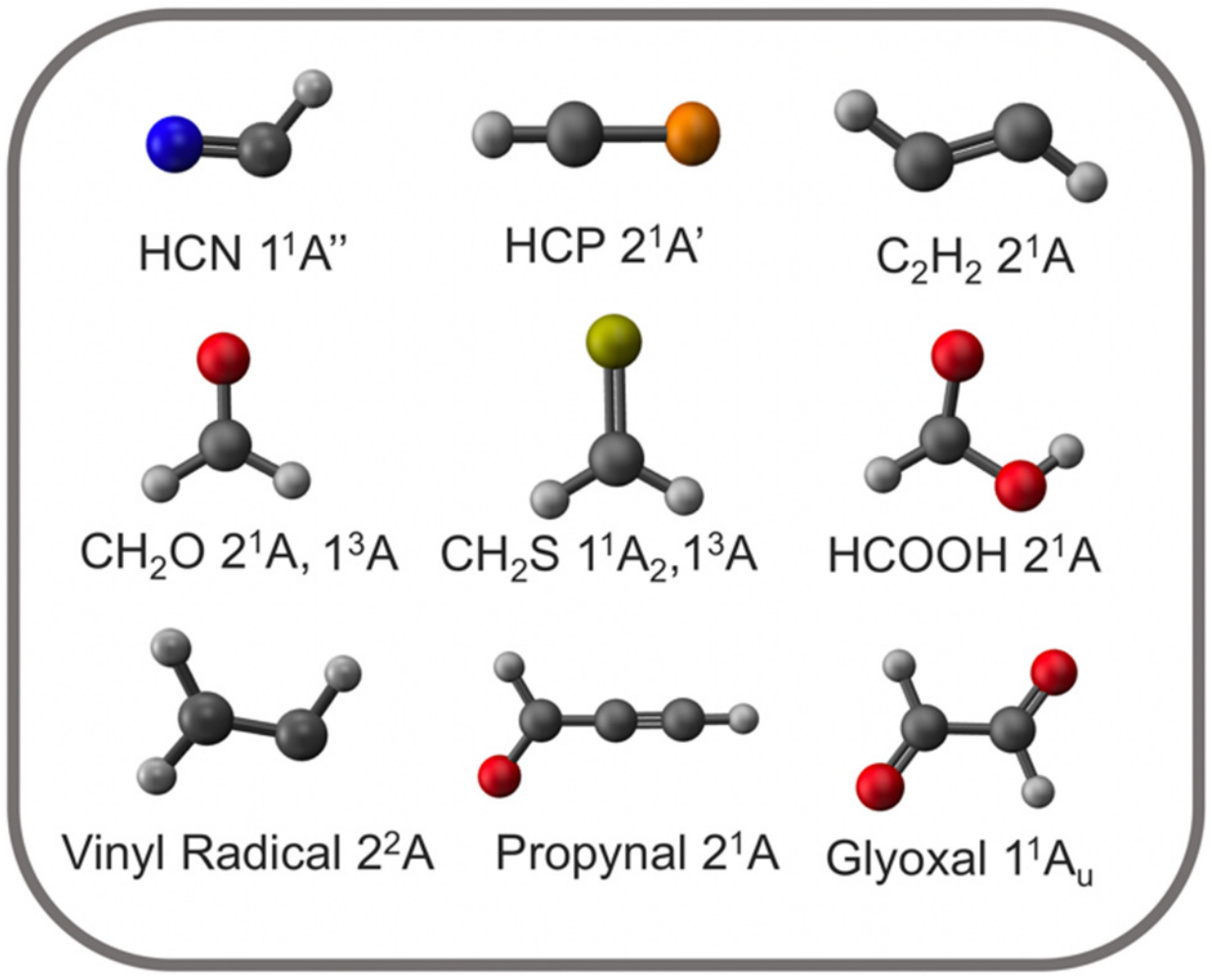

**Figure 3**. Select NEO-TDDFT excited state optimized geometries for the set of nine molecules. These states are used for the calculation of 0-0 adiabatic excitation energies.

The NEO-DFT, NEO-TDDFT, and NEO-TDA calculations, as well as the conventional electronic counterparts, were performed with the B3LYP exc functional[65-66] and the cc-pVTZ electronic basis set.[67] The TDDFT 0-0 adiabatic excitation energies obtained with these choices of exc functional and electronic basis set agree well with previous studies.[68] For NEO-TDDFT calculations, our previous work suggested that larger electronic basis sets, especially those placed on the quantum protons, are crucial for obtaining accurate proton vibrational excitation energies.[35] However, the cc-pVTZ electronic basis set was found to be reasonably converged for NEO-TDDFT calculations of the predominantly electronic excitations, as shown in Table S16. Thus, the NEO calculations were performed with the cc-pVTZ electronic basis set and the PB4-F2a′ (4s3p2d2f) protonic basis set,[69] which was also shown to be reasonably converged, as shown in Table S17. The NEO calculations used either the no-epc treatment, where $E^{\text{epc}} = 0$, or the epc17-2 functional.[25-26] The Cartesian coordinates of all optimized structures are provided in Section S8 of the SI. Adibatic excitation energies for these molecules are also calculated and provided in Table S21.

The 0-0 adiabatic excitation energies for select electronic transitions are presented in Table 1. These predominantly electronic excitations were identified for the NEO calculations by analysis of the transition densities. The NEO and conventional methods produce very similar mean absolute errors (MAEs) and root-mean-square deviations (RMSDs) compared to experimentally measured 0-0 adiabatic excitation energies, although the NEO methods agree slightly better with the experimental values.[70-75] The 0-0 adiabatic excitation energies calculated with NEO-TDDFT and NEO-TDA are somewhat larger than those calculated with their conventional electronic counterparts for all nine molecules except glyoxal. The epc17-2 and no-epc treatments produce very similar results, suggesting that the electron-proton correlation energy is similar in the ground and excited states for predominantly electronic excitations. The TDA and NEO-TDA methods predict slightly larger 0-0 adiabatic excitation energies than the TDDFT and NEO-TDDFT methods, respectively. Although NEO-TDA is known to significantly overestimate proton vibrational frequencies,[33, 35] it produces reasonably accurate 0-0 adiabatic excitation energies for states with predominantly electronic excitation character.

**Table 1**. 0-0 Adiabatic Excitation Energies in eV

| | State | Experiment | TDDFT | NEO-TDDFT epc17-2 | NEO-TDDFT no-epc | TDA | NEO-TDA epc17-2 |
|---|---|---|---|---|---|---|---|
| HCN | $1^1A''$ | 6.48 | 5.99 | 6.13 | 6.11 | 6.02 | 6.16 |
| HCP | $2^1A'$ | 4.31 | 4.31 | 4.38 | 4.38 | 4.32 | 4.38 |
| $C_2H_2$ | $2^1A$ | 5.23 | 4.75 | 4.90 | 4.86 | 4.78 | 4.94 |
| $CH_2O$ | $2^1A$ | 3.49 | 3.62 | 3.63 | 3.64 | 3.66 | 3.68 |
| $CH_2O$ | $1^3A$ | 3.12 | 2.73 | 2.80 | 2.80 | 2.87 | 2.91 |
| $CH_2S$ | $1^1A_2$ | 2.03 | 2.07 | 2.08 | 2.09 | 2.11 | 2.13 |
| $CH_2S$ | $1^3A$ | 1.80 | 1.45 | 1.47 | 1.48 | 1.53 | 1.55 |
| HCOOH | $2^1A$ | 4.64 | 4.79 | 4.87 | 4.86 | 4.82 | 4.90 |
| Vinyl Radical | $2^2A$ | 2.48 | 2.57 | 2.70 | 2.65 | 2.65 | 2.76 |
| Propynal | $2^1A$ | 3.24 | 3.25 | 3.27 | 3.28 | 3.28 | 3.31 |
| Glyoxal | $1^1A_u$ | 2.72 | 2.42 | 2.40 | 2.41 | 2.45 | 2.44 |
| | | MAE | 0.222 | 0.217 | 0.219 | 0.214 | 0.212 |
| | | RMSD | 0.284 | 0.247 | 0.250 | 0.257 | 0.229 |

### 3.2 Excited State Geometry Optimizations of BPOH and $BP(OH)_2$

We performed geometry optimizations on the excited state surfaces corresponding to the HOMO (highest occupied molecular orbital) to LUMO (lowest unoccupied molecular orbital) transition with $\pi\pi^*$ character for the BPOH and $BP(OH)_2$ molecules. In the NEO-TDDFT calculations for these two molecules, the single transferring proton in BPOH and the two transferring protons in $BP(OH)_2$ were treated quantum mechanically. The ωB97X exc functional[76] was used to study the K* and Kt* structures of BPOH due to its reasonable performance compared to resolution-of-identity second-order approximate coupled-cluster (RI-CC2) calculations,[77-78] as shown in Table S18. The B3LYP exc functional was used to study the EE*, EK* and KK* structures in $BP(OH)_2$ because this functional has been previously benchmarked for these structures.[56] Note that the ωB97X functional was not used to study $BP(OH)_2$ because it predicts the KK* structure lower than the EK* structure, as shown in Table S19, which does not agree with the experimental results. The NEO calculations either used the no-epc treatment or the epc17-2 functional. All calculations were performed with the 6-31G(d,p) electronic basis set,[79] and an even-tempered 6s6p6d protonic basis set with exponents ranging from $4\sqrt{2}$ to 32 was used for all NEO calculations, with convergence shown in Table S20.

The geometries shown in Figure 2 were optimized with NEO-DFT and conventional electronic DFT for the ground states and NEO-TDDFT and conventional electronic TDDFT for the excited states. For BPOH, both K* and Kt* stationary points were located with conventional TDDFT and NEO-TDDFT. For $BP(OH)_2$, stationary points corresponding to the planar EE*, EK* and KK* structures were found with conventional TDDFT and NEO-TDDFT. An additional stationary point corresponding to the EK* structure with a slight twist angle of 23°-28° and a lower energy compared to the planar structure by 0.05 eV was also found with NEO-TDDFT (Table S24). Only the twisted EK* structure was observed previously with RI-CC2/TZVP,[56] and only the planar EK* structure was observed with conventional TDDFT/6-31G(d,p). The stationary points on the ground and excited state potential energy surfaces optimized with conventional DFT and TDDFT were confirmed to be minima by analytical Hessian calculations. The excited state NEO-TDDFT Hessian calculations were not performed for these molecules due to the high computational cost of semi-numerical Hessians. The Cartesian coordinates of all optimized structures are provided in Section S8 of the SI.

The differences between the NEO optimized planar structures and the conventionally optimized planar structures are relatively small, as shown in Figure 4, but these differences indicate a strengthening of the intramolecular hydrogen bond upon quantization of the hydrogen nucleus. Specifically, the distance between the transferring proton and the covalently-bonded donor atom is 0.028–0.053 Å longer, and the distance between the transferring proton and the hydrogen-bonded acceptor atom is 0.052–0.115 Å shorter upon proton quantization. Note that the positions of the quantum protons are computed from their expectation values. In addition, the proton donor-acceptor distances are shorter in the NEO optimized structures, with a range of 0.008–0.047 Å, upon proton quantization. The K* structure has a torsion angle of 20.8° and 30.0° when optimized with the NEO-TDDFT method and the conventional TDDFT method, respectively (Table S24). This non-planarity for the K* structure complicates the analysis of the proton donor-acceptor distance but still implicates a stronger intramolecular hydrogen bond. Thus, the ground state and excited state geomeries optimized with the NEO approach exhibit stronger intramolecular hydrogen-bonding interactions, presumably due to the inclusion of proton delocalization and anharmonicity.

Using the optimized geometries, we also calculated the adiabatic excitation energies relative to the ground state energy, as given in Table 2. The ground state geometry optimizations with both conventional DFT and NEO-DFT indicate that the most stable form of both molecules studied is the enol tautomer (E or EE). For BPOH, both the conventional TDDFT and NEO-TDDFT methods predict that the Kt* geometry is lower in energy than the K* geometry, suggesting that the twist is energetically favorable. This finding is consistent with the TICT mechanism leading to fluorescence quenching in BPOH. For $BP(OH)_2$, both methods predict that the planar EK* geometry, where only one proton transfers, is lower in energy than the KK* geometry, where both protons transfer, which is consistent with previous computational and experimental studies.[51, 56] However, a quantitative analysis of the relative stabilities would require larger basis sets and possibly higher levels of electronic structure.

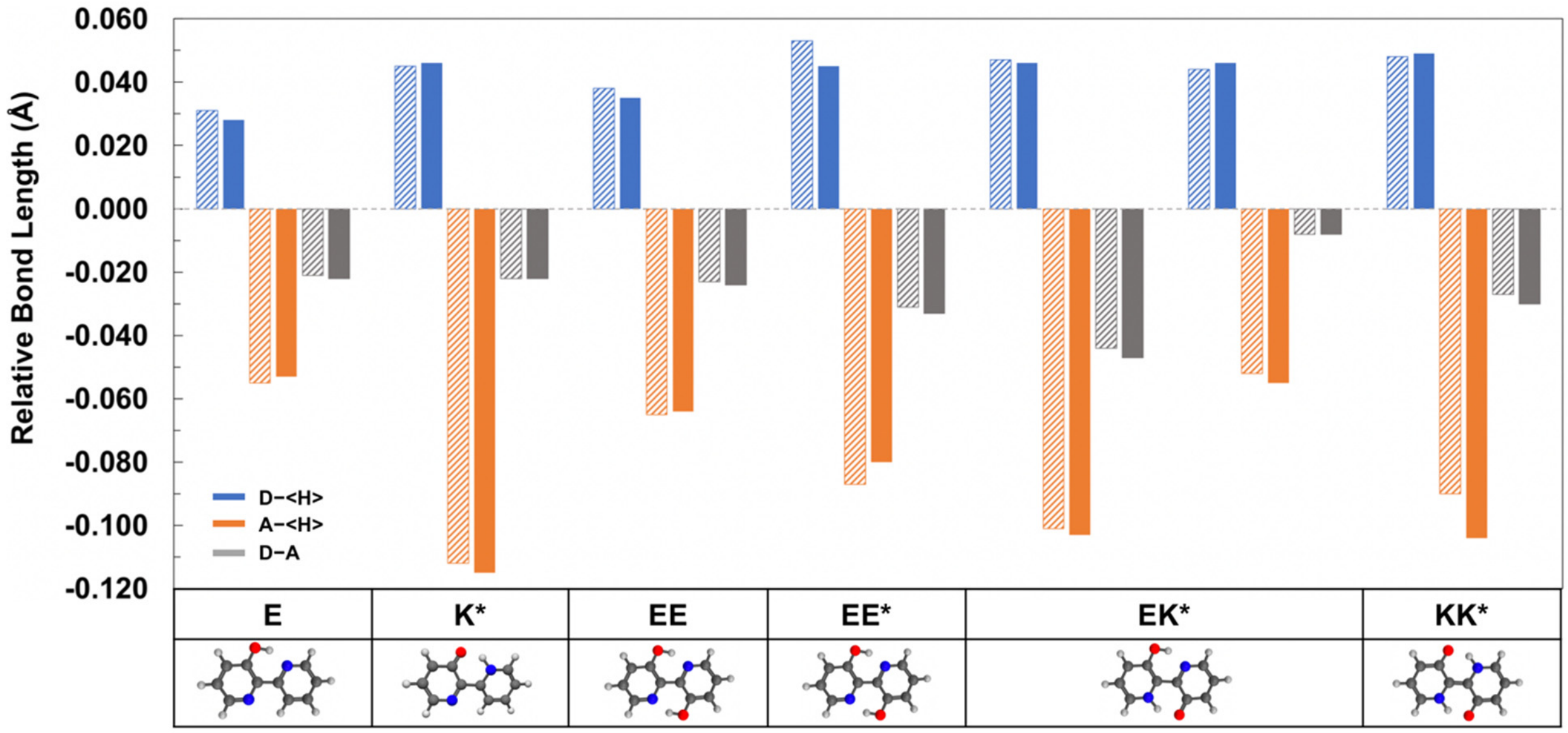

**Figure 4**. Bond lengths at the proton transfer interfaces for BPOH and $BP(OH)_2$ structures obtained from NEO-DFT or NEO-TDDFT geometry optimizations relative to the values obtained for structures obtained from the corresponding conventional DFT or TDDFT geometry optimizations. The notation for the structures is defined in Figure 2, where * indicates an excited state, and E and K denote enol and keto forms, respectively. The lighter, textured bars and the darker, solid bars indicate the NEO calculations with no-epc and the epc17-2 functional, respectively. The colors blue, orange, and gray correspond to the distance between the covalently-bonded donor and the expectation value of the quantum proton coordinate (D–<H>), the distance between the hydrogen-bonded acceptor and the expectation value of the quantum proton coordinate (A–<H>), and the proton donor-acceptor distance (D–A), respectively. Note that the donor is O for the enol form and N for the keto form. For the EK* form, the left and right sets of bars correspond to the enol and keto tautomer interfaces, respectively, within $BP(OH)_2$.

**Table 2**. Adiabatic Excitation Energies (eV) for BPOH and $BP(OH)_2$

| | BPOH | | $BP(OH)_2$ | | |
|---|---|---|---|---|---|
| | K* | Kt* | EE* | EK* | KK* |
| TDDFT | 3.75 | 3.65 | 3.61 | 3.17 | 3.29 |
| TDDFT+FZPE[a] | 3.64 | 3.57 | 3.43 | 3.06 | 3.17 |
| TDDFT+HZPE[b] | 3.74 | 3.64 | 3.54 | 3.19 | 3.29 |
| NEO-TDDFT/no-epc | 3.79 | 3.75 | 3.53 | 3.19 (3.24)[c] | 3.34 |
| NEO-TDDFT/epc17-2 | 3.80 | 3.77 | 3.54 | 3.20 (3.25)[c] | 3.35 |
| NEO-TDDFT/epc17-2/conv[d] | 3.81 | 3.77 | 3.55 | 3.25 | 3.36 |

[a]Adiabatic excitation energy includes the harmonic ZPE difference for all nuclei.
[b]Adiabatic excitation energy includes the harmonic ZPE difference for only the transferring hydrogen(s) by setting the other masses to large values for the normal mode analysis.
[c]Two stationary points were located for EK*. The planar structure is higher in energy than the slightly twisted geometry, which has a twist angle of 22.9°-28.3°, and the energy associated with it is given in parentheses.
[d]Adiabatic excitation energy calculated with NEO-TDDFT/epc17-2 at the conventional TDDFT and DFT excited state and ground state optimized geometries.

According to Table 2, the adiabatic excitation energies computed with NEO-TDDFT are 0.02 – 0.12 eV higher than those computed with conventional TDDFT, except for the EE* structure. Because the NEO-TDDFT approach inherently includes the ZPE associated with the quantum protons, its adiabatic excitation energy includes the difference between the protonic ZPE in the excited state and the ground state. Thus, to enable a direct comparison, we computed both the full ZPE contribution from all nuclei and only the protonic ZPE contribution to the conventional TDDFT excitation energy, denoted FZPE and HZPE, respectively, within the harmonic approximation. For the HZPE calculations, all nuclei other than the quantum protons were assigned large masses in the normal mode analysis to eliminate their ZPE contributions. The harmonic FZPE treatment predicts the full ZPE contribution to be lower in the excited state than in the ground state, thereby lowering the adiabatic excitation energies. In contrast, the protonic ZPE is similar for both the ground and excited states for all structures except the EE* structure, where the harmonic HZPE treatment leads to better agreement with the NEO-TDDFT results.

The results in Table 2 indicate that the NEO-TDDFT adiabatic excitation energies obtained with no-epc and with the epc17-2 functional differ by less than ~0.02 eV. These results suggest that the electron-proton correlation energy is similar in the excited and ground electronic states at their corresponding optimized geometries and therefore cancels out. However, the epc17-2 functional significantly improves the absolute protonic ZPE for the ground and excited states compared to the no-epc treatment (Table S22). Because the protonic ZPEs are nearly the same in the ground and excited states for both the no-epc and epc17-2 treatments, these ZPEs cancel out in both cases when calculating the adiabatic excitation energies.

To examine the effects of differences in the NEO and conventional optimized geometries, we performed NEO-DFT and NEO-TDDFT calculations at the geometries optimized with conventional DFT and TDDFT, respectively, denoted NEO-TDDFT/epc17-2/conv in Table 2. However, the geometric effect on the NEO-TDDFT adiabatic excitation energies was found to be only ~0.01eV. The differences between the NEO-TDDFT results and the conventional TDDFT with protonic ZPE (TDDFT+HZPE) results may be due to the anharmonic effects of the quantum protons and the vibronic mixing between electronic and proton vibrational states that are included inherently in NEO calculations but are absent in conventional calculations. Limitations of the epc17-2 functional may also impact quantitative accuracy.

## 4. Conclusions

In this paper, we present the derivation of the NEO-TDDFT and NEO-TDA analytical gradients and the corresponding programmable equations. By treating the protons quantum mechanically within the NEO framework, protonic ZPE, density delocalization, and anharmonicity are included inherently in energy calculations and geometry optimizations. We used these analytical gradients to compute the 0-0 adiabatic excitation energies for a set of nine small molecules and found that the NEO methods lead to slightly better agreement with experimental data compared to conventional electronic methods. We also used these methods to perform excited state geometry optimizations of BPOH and $BP(OH)_2$, which are prototypes for ESIPT. The geometries optimized with the NEO methods are similar to those optimized with conventional electronic methods except that they exhibit stronger intramolecular hydrogen bonds for the planar geometries. Moreover, an additional slightly twisted EK* stationary point was located with the NEO-TDDFT method but not with the conventional TDDFT method for the same electronic basis set. The relative stabilities of different excited state structures are similar for the NEO-TDDFT and conventional electronic TDDFT calculations.

The analytical NEO-TDDFT gradients and the capability for excited state geometry optimizations demonstrated by these applications provide the foundation for investigating interesting and challenging vibronically nonadiabatic systems. In addition to static excited state properties, NEO-TDDFT analytical gradients also serve as the basis for both adiabatic and nonadiabatic dynamics on the NEO vibronic surfaces. Although the NEO Ehrenfest nonadiabatic dynamics approach has been developed in conjunction with NEO real-time TDDFT,[80-82] nonadiabatic dynamics approaches such as surface hopping[83] require the use of adiabatic excited state potential energy surfaces and the gradients of these surfaces. The linear-response NEO-TDDFT method with analytical gradients developed herein will be essential for these types of simulations. Thus, this work is an important step toward NEO simulations of fundamental processes such as photoinduced proton transfer and proton-coupled electron transfer.

## Supporting Information

The Supporting Information is available free of charge on the ACS Publications website. Derivations of the Z vector equations for NEO-TDDFT and NEO-TDA analytical gradients, benchmarking analytical gradients against numerical gradients, equations for NEO-DFT analytical Hessian, benchmarking electronic exchange-correlation functionals, electronic and protonic basis set convergence tests, adiabatic excitation energies for set of nine molecules, protonic zero-point energies, geometric analysis of ESIPT systems, and coordinates of optimized geometries for molecules studied.

**Acknowledgements**

This work was supported by the National Science Foundation Grant No. CHE-1954348. P.E.S. is supported by a National Science Foundation Graduate Research Fellowship Grant No. DGE-1752134. The authors thank Dr. Yang Yang, Dr. Tanner Culpitt, Dr. Evgeny Epifanovsky, Dr. Xintian Feng, Dr. Qi Yu, Ben Rousseau, Dr. Christopher Malbon, Dr. John Tully, and Mathew Chow for useful discussions.

**References**

(1) Wang, J.; Durbeej, B. How accurate are TD-DFT excited-state geometries compared to DFT ground-state geometries? *J. Comput. Chem.* **2020,** *41*, 1718-1729.
(2) Loos, P.-F.; Jacquemin, D. Evaluating 0–0 Energies with Theoretical Tools: A Short Review. *ChemPhotoChem* **2019,** *3*, 684-696.
(3) Curchod, B. F. E.; Rothlisberger, U.; Tavernelli, I. Trajectory-Based Nonadiabatic Dynamics with Time-Dependent Density Functional Theory. *ChemPhysChem* **2013,** *14*, 1314-1340.
(4) Crespo-Otero, R.; Barbatti, M. Recent Advances and Perspectives on Nonadiabatic Mixed Quantum–Classical Dynamics. *Chem. Rev.* **2018,** *118*, 7026-7068.
(5) Rappoport, D.; Furche, F., Excited States and Photochemistry. In *Time-Dependent Density Functional Theory*, Marques, M. A. L.; Ullrich, C. A.; Nogueira, F.; Rubio, A.; Burke, K.; Gross, E. K. U., Eds. Springer Berlin Heidelberg: Berlin, Heidelberg, 2006; pp 337-354.
(6) Tapavicza, E.; Tavernelli, I.; Rothlisberger, U. Trajectory Surface Hopping within Linear Response Time-Dependent Density-Functional Theory. *Phys. Rev. Lett.* **2007,** *98*, 023001.
(7) Tozer, D. J.; Amos, R. D.; Handy, N. C.; Roos, B. O.; Serrano-Andres, L. Does density functional theory contribute to the understanding of excited states of unsaturated organic compounds? *Mol. Phys.* **1999,** *97*, 859-868.

(8) Casida, M. E.; Gutierrez, F.; Guan, J.; Gadea, F.-X.; Salahub, D.; Daudey, J.-P. Charge-transfer correction for improved time-dependent local density approximation excited-state potential energy curves: Analysis within the two-level model with illustration for H2 and LiH. *J. Chem. Phys.* **2000,** *113*, 7062-7071.
(9) Grimme, S.; Parac, M. Substantial Errors from Time-Dependent Density Functional Theory for the Calculation of Excited States of Large π Systems. *ChemPhysChem* **2003,** *4*, 292-295.
(10) Dreuw, A.; Weisman, J. L.; Head-Gordon, M. Long-range charge-transfer excited states in time-dependent density functional theory require non-local exchange. *J. Chem. Phys.* **2003,** *119*, 2943-2946.
(11) Dreuw, A.; Head-Gordon, M. Single-Reference ab Initio Methods for the Calculation of Excited States of Large Molecules. *Chem. Rev.* **2005,** *105*, 4009-4037.
(12) Fuks, J. I. Time-dependent density functional theory for charge-transfer dynamics: review of the causes of failure and success*. *Eur. Phys. J. B* **2016,** *89*, 236.
(13) Maitra, N. T. Charge transfer in time-dependent density functional theory. *J. Phys. Condens. Matter* **2017,** *29*, 423001.
(14) Levine, B. G.; Ko, C.; Quenneville, J.; Martínez, T. J. Conical intersections and double excitations in time-dependent density functional theory. *Mol. Phys.* **2006,** *104*, 1039-1051.
(15) Tsuneda, T.; Hirao, K. Long-range correction for density functional theory. *WIREs Computational Molecular Science* **2014,** *4*, 375-390.
(16) Hirata, S.; Head-Gordon, M. Time-dependent density functional theory within the Tamm–Dancoff approximation. *Chem. Phys. Lett.* **1999,** *314*, 291-299.
(17) Tapavicza, E.; Tavernelli, I.; Rothlisberger, U.; Filippi, C.; Casida, M. E. Mixed time-dependent density-functional theory/classical trajectory surface hopping study of oxirane photochemistry. *J. Chem. Phys.* **2008,** *129*, 124108.
(18) Autschbach, J. Charge-Transfer Excitations and Time-Dependent Density Functional Theory: Problems and Some Proposed Solutions. *ChemPhysChem* **2009,** *10*, 1757-1760.
(19) Herbert, J. M.; Zhang, X.; Morrison, A. F.; Liu, J. Beyond Time-Dependent Density Functional Theory Using Only Single Excitations: Methods for Computational Studies of Excited States in Complex Systems. *Acc. Chem. Res.* **2016,** *49*, 931-941.
(20) Morrone, J. A.; Car, R. Nuclear Quantum Effects in Water. *Phys. Rev. Lett.* **2008,** *101*, 017801.
(21) Hammes-Schiffer, S. Proton-Coupled Electron Transfer: Moving Together and Charging Forward. *J. Am. Chem. Soc.* **2015,** *137*, 8860-8871.
(22) Webb, S. P.; Iordanov, T.; Hammes-Schiffer, S. Multiconfigurational nuclear-electronic orbital approach: Incorporation of nuclear quantum effects in electronic structure calculations. *J. Chem. Phys.* **2002,** *117*, 4106-4118.
(23) Pavošević, F.; Culpitt, T.; Hammes-Schiffer, S. Multicomponent Quantum Chemistry: Integrating Electronic and Nuclear Quantum Effects via the Nuclear–Electronic Orbital Method. *Chem. Rev.* **2020,** *120*, 4222-4253.
(24) Pak, M. V.; Chakraborty, A.; Hammes-Schiffer, S. Density Functional Theory Treatment of Electron Correlation in the Nuclear−Electronic Orbital Approach. *J. Phys. Chem. A* **2007,** *111*, 4522-4526.
(25) Yang, Y.; Brorsen, K. R.; Culpitt, T.; Pak, M. V.; Hammes-Schiffer, S. Development of a practical multicomponent density functional for electron-proton correlation to produce accurate proton densities. *J. Chem. Phys.* **2017,** *147*, 114113.
(26) Brorsen, K. R.; Yang, Y.; Hammes-Schiffer, S. Multicomponent Density Functional Theory: Impact of Nuclear Quantum Effects on Proton Affinities and Geometries. *J. Phys. Chem. Lett.* **2017,** *8*, 3488-3493.
(27) Brorsen, K. R.; Schneider, P. E.; Hammes-Schiffer, S. Alternative forms and transferability of electron-proton correlation functionals in nuclear-electronic orbital density functional theory. *J. Chem. Phys.* **2018,** *149*, 044110.

(28) Tao, Z.; Yang, Y.; Hammes-Schiffer, S. Multicomponent density functional theory: Including the density gradient in the electron-proton correlation functional for hydrogen and deuterium. *J. Chem. Phys.* **2019,** *151*, 124102.
(29) Pavošević, F.; Culpitt, T.; Hammes-Schiffer, S. Multicomponent Coupled Cluster Singles and Doubles Theory within the Nuclear-Electronic Orbital Framework. *J. Chem. Theory Comput.* **2019,** *15*, 338-347.
(30) Pavošević, F.; Rousseau, B. J. G.; Hammes-Schiffer, S. Multicomponent Orbital-Optimized Perturbation Theory Methods: Approaching Coupled Cluster Accuracy at Lower Cost. *J. Phys. Chem. Lett.* **2020,** *11*, 1578-1583.
(31) Pavošević, F.; Tao, Z.; Hammes-Schiffer, S. Multicomponent Coupled Cluster Singles and Doubles with Density Fitting: Protonated Water Tetramers with Quantized Protons. *J. Phys. Chem. Lett* **2021**, 1631-1637.
(32) Schneider, P. E.; Tao, Z.; Pavošević, F.; Epifanovsky, E.; Feng, X.; Hammes-Schiffer, S. Transition states, reaction paths, and thermochemistry using the nuclear–electronic orbital analytic Hessian. *J. Chem. Phys.* **2021,** *154*, 054108.
(33) Yang, Y.; Culpitt, T.; Hammes-Schiffer, S. Multicomponent Time-Dependent Density Functional Theory: Proton and Electron Excitation Energies. *J. Phys. Chem. Lett.* **2018,** *9*, 1765-1770.
(34) Pavošević, F.; Hammes-Schiffer, S. Multicomponent equation-of-motion coupled cluster singles and doubles: Theory and calculation of excitation energies for positronium hydride. *J. Chem. Phys.* **2019,** *150*, 161102.
(35) Culpitt, T.; Yang, Y.; Pavošević, F.; Tao, Z.; Hammes-Schiffer, S. Enhancing the applicability of multicomponent time-dependent density functional theory. *J. Chem. Phys.* **2019,** *150*, 201101.
(36) Pavošević, F.; Tao, Z.; Culpitt, T.; Zhao, L.; Li, X.; Hammes-Schiffer, S. Frequency and Time Domain Nuclear–Electronic Orbital Equation-of-Motion Coupled Cluster Methods: Combination Bands and Electronic–Protonic Double Excitations. *J. Phys. Chem. Lett.* **2020,** *11*, 6435-6442.
(37) Capitani, J. F.; Nalewajski, R. F.; Parr, R. G. Non-Born–Oppenheimer density functional theory of molecular systems. *J. Chem. Phys.* **1982,** *76*, 568-573.
(38) Shigeta, Y.; Takahashi, H.; Yamanaka, S.; Mitani, M.; Nagao, H.; Yamaguchi, K. Density functional theory without the Born–Oppenheimer approximation and its application. *Int. J. Quantum Chem.* **1998,** *70*, 659-669.
(39) Gidopoulos, N. Kohn-Sham equations for multicomponent systems: The exchange and correlation energy functional. *Phys. Rev. B* **1998,** *57*, 2146-2152.
(40) Kreibich, T.; Gross, E. K. U. Multicomponent Density-Functional Theory for Electrons and Nuclei. *Phys. Rev. Lett.* **2001,** *86*, 2984-2987.
(41) Nakai, H.; Sodeyama, K. Many-body effects in nonadiabatic molecular theory for simultaneous determination of nuclear and electronic wave functions: Ab initio NOMO/MBPT and CC methods. *J. Chem. Phys.* **2003,** *118*, 1119-1127.
(42) Bubin, S.; Cafiero, M.; Adamowicz, L., Non-Born–Oppenheimer Variational Calculations of Atoms and Molecules with Explicitly Correlated Gaussian Basis Functions. In *Advances in Chemical Physics*, 2005; pp 377-475.
(43) van Leeuwen, R.; Gross, E. K. U., Multicomponent Density-Functional Theory. In *Time-Dependent Density Functional Theory*, Marques, M. A. L.; Ullrich, C. A.; Nogueira, F.; Rubio, A.; Burke, K.; Gross, E. K. U., Eds. Springer Berlin Heidelberg: Berlin, Heidelberg, 2006; pp 93-106.
(44) Ishimoto, T.; Tachikawa, M.; Nagashima, U. Review of multicomponent molecular orbital method for direct treatment of nuclear quantum effect. *Int. J. Quantum Chem.* **2009,** *109*, 2677-2694.
(45) Chakraborty, A.; Pak, M. V.; Hammes-Schiffer, S. Properties of the exact universal functional in multicomponent density functional theory. *J. Chem. Phys.* **2009,** *131*, 124115.
(46) Kwon, J. E.; Park, S. Y. Advanced Organic Optoelectronic Materials: Harnessing Excited-State Intramolecular Proton Transfer (ESIPT) Process. *Adv. Mater.* **2011,** *23*, 3615-3642.
(47) Bulska, H.; Grabowska, A.; Grabowski, Z. R. Single and double proton transfer in excited hydroxy derivatives of bipyridyl. *J. Lumin.* **1986,** *35*, 189-197.

(48) Tokumura, K.; Oyama, O.; Mukaihata, H.; Itoh, M. Rotational Isomerization of Phototautomer Produced in the Excited-State Proton Transfer of 2,2‘-Bipyridin-3-ol. *J. Phys. Chem. A* **1997,** *101*, 1419-1421.
(49) Vollmer, F.; Rettig, W. Fluorescence loss mechanism due to large-amplitude motions in derivatives of 2,2′-bipyridyl exhibiting excited-state intramolecular proton transfer and perspectives of luminescence solar concentrators. *J. Photochem. Photobiol. A Chem.* **1996,** *95*, 143-155.
(50) Kim, S.; Seo, J.; Park, S. Y. Torsion-induced fluorescence quenching in excited-state intramolecular proton transfer (ESIPT) dyes. *J. Photochem. Photobiol. A Chem.* **2007,** *191*, 19-24.
(51) Zhang, H.; van der Meulen, P.; Glasbeek, M. Ultrafast single and double proton transfer in photo-excited [2,2′-bipyridyl]-3,3′-diol. *Chem. Phys. Lett.* **1996,** *253*, 97-102.
(52) Neuwahl, F. V. R.; Foggi, P.; Brown, R. G. Sub-picosecond and picosecond dynamics in the S1 state of [2,2′-bipyridyl]-3,3′-diol investigated by UV–visible transient absorption spectroscopy. *Chem. Phys. Lett.* **2000,** *319*, 157-163.
(53) Barone, V.; Adamo, C. A theoretical study of proton transfer in [2,2′-bipyridyl]-3,3′-diol. *Chem. Phys. Lett.* **1995,** *241*, 1-6.
(54) Sobolewski, A. L.; Adamowicz, L. Double-proton-transfer in [2,2′-bipyridine]-3,3′-diol: an ab initio study. *Chem. Phys. Lett.* **1996,** *252*, 33-41.
(55) Gelabert, R.; Moreno, M.; Lluch, J. M. Quantum Dynamics Study of the Excited-State Double-Proton Transfer in 2,2′-Bipyridyl-3,3′-diol. *ChemPhysChem* **2004,** *5*, 1372-1378.
(56) Plasser, F.; Barbatti, M.; Aquino, A. J. A.; Lischka, H. Excited-State Diproton Transfer in [2,2′-Bipyridyl]-3,3′-diol: the Mechanism Is Sequential, Not Concerted. *J. Phys. Chem. A* **2009,** *113*, 8490-8499.
(57) Ortiz-Sánchez, J. M.; Gelabert, R.; Moreno, M.; Lluch, J. M.; Anglada, J. M.; Bofill, J. M. Bipyridyl Derivatives as Photomemory Devices: A Comparative Electronic-Structure Study. *Chemistry – A European Journal* **2010,** *16*, 6693-6703.
(58) Auer, B.; Hammes-Schiffer, S. Localized Hartree product treatment of multiple protons in the nuclear-electronic orbital framework. *J. Chem. Phys.* **2010,** *132*, 084110.
(59) Colle, R.; Salvetti, O. Approximate calculation of the correlation energy for the closed shells. *Theor. Chim. Acta* **1975,** *37*, 329-334.
(60) Furche, F.; Ahlrichs, R. Adiabatic time-dependent density functional methods for excited state properties. *J. Chem. Phys.* **2002,** *117*, 7433-7447.
(61) Shao, Y.; Gan, Z.; Epifanovsky, E.; Gilbert, A. T. B.; Wormit, M.; Kussmann, J.; Lange, A. W.; Behn, A.; Deng, J.; Feng, X.; Ghosh, D.; Goldey, M.; Horn, P. R.; Jacobson, L. D.; Kaliman, I.; Khaliullin, R. Z.; Kuś, T.; Landau, A.; Liu, J.; Proynov, E. I.; Rhee, Y. M.; Richard, R. M.; Rohrdanz, M. A.; Steele, R. P.; Sundstrom, E. J.; Woodcock, H. L.; Zimmerman, P. M.; Zuev, D.; Albrecht, B.; Alguire, E.; Austin, B.; Beran, G. J. O.; Bernard, Y. A.; Berquist, E.; Brandhorst, K.; Bravaya, K. B.; Brown, S. T.; Casanova, D.; Chang, C.-M.; Chen, Y.; Chien, S. H.; Closser, K. D.; Crittenden, D. L.; Diedenhofen, M.; DiStasio, R. A.; Do, H.; Dutoi, A. D.; Edgar, R. G.; Fatehi, S.; Fusti-Molnar, L.; Ghysels, A.; Golubeva-Zadorozhnaya, A.; Gomes, J.; Hanson-Heine, M. W. D.; Harbach, P. H. P.; Hauser, A. W.; Hohenstein, E. G.; Holden, Z. C.; Jagau, T.-C.; Ji, H.; Kaduk, B.; Khistyaev, K.; Kim, J.; Kim, J.; King, R. A.; Klunzinger, P.; Kosenkov, D.; Kowalczyk, T.; Krauter, C. M.; Lao, K. U.; Laurent, A. D.; Lawler, K. V.; Levchenko, S. V.; Lin, C. Y.; Liu, F.; Livshits, E.; Lochan, R. C.; Luenser, A.; Manohar, P.; Manzer, S. F.; Mao, S.-P.; Mardirossian, N.; Marenich, A. V.; Maurer, S. A.; Mayhall, N. J.; Neuscamman, E.; Oana, C. M.; Olivares-Amaya, R.; O'Neill, D. P.; Parkhill, J. A.; Perrine, T. M.; Peverati, R.; Prociuk, A.; Rehn, D. R.; Rosta, E.; Russ, N. J.; Sharada, S. M.; Sharma, S.; Small, D. W.; Sodt, A.; Stein, T.; Stück, D.; Su, Y.-C.; Thom, A. J. W.; Tsuchimochi, T.; Vanovschi, V.; Vogt, L.; Vydrov, O.; Wang, T.; Watson, M. A.; Wenzel, J.; White, A.; Williams, C. F.; Yang, J.; Yeganeh, S.; Yost, S. R.; You, Z.-Q.; Zhang, I. Y.; Zhang, X.; Zhao, Y.; Brooks, B. R.; Chan, G. K. L.; Chipman, D. M.; Cramer, C. J.; Goddard, W. A.; Gordon, M. S.; Hehre, W. J.; Klamt, A.; Schaefer, H. F.; Schmidt, M. W.; Sherrill, C. D.; Truhlar, D. G.; Warshel, A.; Xu, X.; Aspuru-Guzik, A.; Baer, R.; Bell, A. T.; Besley, N. A.; Chai, J.-D.; Dreuw, A.; Dunietz, B. D.; Furlani, T. R.; Gwaltney, S. R.; Hsu, C.-P.; Jung, Y.; Kong,

J.; Lambrecht, D. S.; Liang, W.; Ochsenfeld, C.; Rassolov, V. A.; Slipchenko, L. V.; Subotnik, J. E.; Van Voorhis, T.; Herbert, J. M.; Krylov, A. I.; Gill, P. M. W.; Head-Gordon, M. Advances in molecular quantum chemistry contained in the Q-Chem 4 program package. *Mol. Phys.* **2015,** *113*, 184-215.
(62) Stratmann, R. E.; Scuseria, G. E.; Frisch, M. J. An efficient implementation of time-dependent density-functional theory for the calculation of excitation energies of large molecules. *J. Chem. Phys.* **1998,** *109*, 8218-8224.
(63) Davidson, E. R. The iterative calculation of a few of the lowest eigenvalues and corresponding eigenvectors of large real-symmetric matrices. *J. Comput. Phys.* **1975,** *17*, 87-94.
(64) O'Leary, D. P. The block conjugate gradient algorithm and related methods. *Linear Algebra and its Applications* **1980,** *29*, 293-322.
(65) Lee, C.; Yang, W.; Parr, R. G. Development of the Colle-Salvetti correlation-energy formula into a functional of the electron density. *Phys. Rev. B* **1988,** *37*, 785-789.
(66) Becke, A. D. Density-functional thermochemistry. III. The role of exact exchange. *J. Chem. Phys.* **1993,** *98*, 5648-5652.
(67) Dunning, T. H., Jr. Gaussian basis sets for use in correlated molecular calculations. I. The atoms boron through neon and hydrogen. *J. Chem. Phys.* **1989,** *90*, 1007-1023.
(68) Send, R.; Kühn, M.; Furche, F. Assessing Excited State Methods by Adiabatic Excitation Energies. *J. Chem. Theory Comput.* **2011,** *7*, 2376-2386.
(69) Yu, Q.; Pavošević, F.; Hammes-Schiffer, S. Development of nuclear basis sets for multicomponent quantum chemistry methods. *J. Chem. Phys.* **2020,** *152*, 244123.
(70) Herzberg, G., *Molecular Spectra and Molecular Structure*. Van Nostrand and Reinhold: New York, 1966; Vol. ii.
(71) Brand, J. C. D.; Chan, W. H.; Liu, D. S.; Callomon, J. H.; Watson, J. K. G. The 3820 Å band system of propynal: Rotational analysis of the 0-0 band. *J. Mol. Spectrosc.* **1974,** *50*, 304-309.
(72) D J Clouthier, a.; Ramsay, D. A. The Spectroscopy of Formaldehyde and Thioformaldehyde. *Annu. Rev. Phys. Chem.* **1983,** *34*, 31-58.
(73) Ioannoni, F.; Moule, D. C.; Clouthier, D. J. Laser spectroscopic and quantum chemical studies of the lowest excited states of formic acid. *J. Phys. Chem.* **1990,** *94*, 2290-2294.
(74) Humphrey, S. J.; Pratt, D. W. High resolution S1←S0 fluorescence excitation spectra of hydroquinone. Distinguishing the cis and trans rotamers by their nuclear spin statistical weights. *J. Chem. Phys.* **1993,** *99*, 5078-5086.
(75) Pushkarsky, M. B.; Mann, A. M.; Yeston, J. S.; Moore, C. B. Electronic spectroscopy of jet-cooled vinyl radical. *J. Chem. Phys.* **2001,** *115*, 10738-10744.
(76) Chai, J.-D.; Head-Gordon, M. Systematic optimization of long-range corrected hybrid density functionals. *J. Chem. Phys.* **2008,** *128*, 084106.
(77) Christiansen, O.; Koch, H.; Jørgensen, P. The second-order approximate coupled cluster singles and doubles model CC2. *Chem. Phys. Lett.* **1995,** *243*, 409-418.
(78) Köhn, A.; Hättig, C. Analytic gradients for excited states in the coupled-cluster model CC2 employing the resolution-of-the-identity approximation. *J. Chem. Phys.* **2003,** *119*, 5021-5036.
(79) Francl, M. M.; Pietro, W. J.; Hehre, W. J.; Binkley, J. S.; Gordon, M. S.; DeFrees, D. J.; Pople, J. A. Self-consistent molecular orbital methods. XXIII. A polarization-type basis set for second-row elements. *J. Chem. Phys.* **1982,** *77*, 3654-3665.
(80) Zhao, L.; Tao, Z.; Pavošević, F.; Wildman, A.; Hammes-Schiffer, S.; Li, X. Real-Time Time-Dependent Nuclear−Electronic Orbital Approach: Dynamics beyond the Born–Oppenheimer Approximation. *J. Phys. Chem. Lett.* **2020,** *11*, 4052-4058.
(81) Zhao, L.; Wildman, A.; Tao, Z.; Schneider, P.; Hammes-Schiffer, S.; Li, X. Nuclear–electronic orbital Ehrenfest dynamics. *J. Chem. Phys.* **2020,** *153*, 224111.
(82) Zhao, L.; Wildman, A.; Pavošević, F.; Tully, J. C.; Hammes-Schiffer, S.; Li, X. Excited State Intramolecular Proton Transfer with Nuclear-Electronic Orbital Ehrenfest Dynamics. *J. Phys. Chem. Lett.* **2021**, 3497-3502.
(83) Tully, J. C. Molecular dynamics with electronic transitions. *J. Chem. Phys.* **1990,** *93*, 1061-1071.

**TOC Graphic**

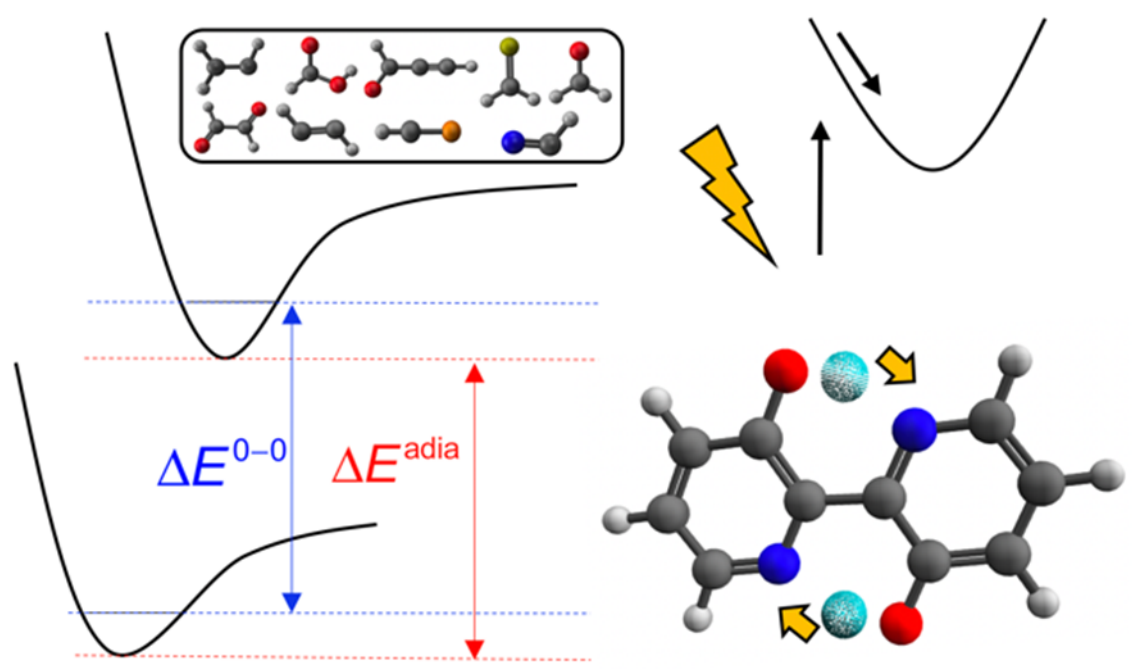